\documentclass[12pt,a4,epsf]{article}
\usepackage{graphicx,showlabels}
\usepackage{latexsym}
\usepackage{natbib}
\usepackage{epsfig}
\usepackage{bm}
\usepackage{amsmath}
\usepackage{amssymb}
\usepackage{color}
\usepackage{amsthm,ctable}
\usepackage{amsthm,hyperref}

\RequirePackage[mathlines, displaymath]{lineno}      

\newtheorem{theo}{\sc Theorem}

\newtheorem{coll}{\sc Corollary}[section]

\newcommand{\be}{\begin{equation}}
\newcommand{\ee}{\end{equation}}
\newcommand{\bea}{\begin{eqnarray}}
\newcommand{\eea}{\end{eqnarray}}
\newcommand{\beas}{\begin{eqnarray*}}
\newcommand{\eeas}{\end{eqnarray*}}

\newtheorem{remark}{{\bf Remark}}

\makeatletter      
\@addtoreset{equation}{section}
\makeatother       

\long \def\@makecation#1#2{ \vskip 10 pt
\setbox\@tempboxa\hbox{#1:#2} \ifdim \wd\@tempboxa >\hsize
\unhbox\@tempboxa\par \else \hbox
to\hsize{\hfil\box\@temboxa\hfil} \fi}

\def\Xi{X^{(i)}}

\begin{document}


\title{ Variable Selection and Estimation for Partially Linear Single-index Models with Longitudinal Data}

\author{Gaorong Li$^{a}$,~  Peng Lai$^{b}$ ~ and~ Heng
Lian$^{c}$\thanks{Heng Lian  is the corresponding author. E-mail:
henglian@ntu.edu.sg.}\\
 \\
\begin{tabular}{l}
{\small\it $^{a}$College of Applied Sciences, Beijing University of Technology, Beijing 100124, China }\\
{\small\it $^{b}$School of Mathematics and Statistics, Nanjing University of Information Science }\\
 {\small\it\hskip4cm  $\&$ Technology, Nanjing 210044, China}\\
{\small\it $^{c}$Division of Mathematical Sciences, SPMS, Nanyang Technological University, Singapore}\\
\end{tabular}
}
\date{}
\maketitle

\begin{quote}

\begin{abstract}

In this paper, we consider the partially linear single-index models with longitudinal data. To deal with the variable selection problem in this context, we propose a penalized procedure combined with two bias correction methods, resulting in the bias-corrected generalized estimating equation (GEE) and the bias-corrected quadratic inference function (QIF), which can take into account the correlations. Asymptotic properties of these methods are demonstrated. We also evaluate the finite sample performance of the proposed methods via Monte Carlo simulation studies and a real data analysis.

\end{abstract}


\noindent {\footnotesize {\it AMS2000 subject classifications}:
primary 62J05; secondary 62J07}
\end{quote}



\newpage
\baselineskip=18pt

\section{Introduction}\label{sec1}

Longitudinal/clustered data modeling is often used in experiments that are designed such that responses on the same experimental units are observed repeatedly. Experiments of this type have extensive applications in many fields, including epidemiology, econometrics, medicine, life and social sciences.  Let $\{(Y_{ij},X_{ij},Z_{ij})_{1\leq i\leq n,1\leq j\leq m_{i}}\}$ be the $j$th observation for the $i$th subject or experimental unit, where $Y_{ij}$ is the response variable associated with explanatory variables $(X_{ij},Z_{ij})\in \mathbb{R}^{p}\times \mathbb{R}^{q}$. Throughout this paper we assume that $n$ increases to push up the total sample size $N=\sum_{i=1}^{n}m_{i}$, while $\{m_{i}\}$ is a bounded sequence of positive integers. This means that $n$ and $N$ have the same order. The partially linear single-index model for longitudinal data takes the form
\begin{equation}\label{PLSIM}
Y_{ij}=g(X_{ij}^{{\rm T}}\bm\beta_0)+Z_{ij}^{{\rm T}}\bm\theta_0+e_{ij},\quad i=1,\ldots,n,\quad j=1,\ldots,m_{i},
\end{equation}
where  $(\bm\beta_0,\bm\theta_0)$ is an unknown vector in $\mathbb{R}^{p}\times \mathbb{R}^{q}$  with $\|\bm\beta_0\|=1~$(where $\|\cdot\|$ denotes the Euclidean norm), $g(\cdot)$ is an unknown univariate link function, ${\bm e}_{i}=(e_{i1},e_{i2},\ldots,e_{im_{i}})^{\rm T}$ is the random error vector of the $i$th subject, and $\{{\bm e}_{i},i=1,\ldots,n\}$ are mutually independent with $E({\bm e}_{i}|{\bm X}_{i},{\bm Z}_{i})=0$ and ${\rm{Var}}({\bm e}_{i})=\Sigma_{i}$. The constraint $\|\bm\beta_0\|=1$ is for the identifiability of $\bm\beta_0$.

Model (\ref{PLSIM}) has been  studied by many authors, for example \cite{li2010empirical}, \cite{lai2013QIF} and \cite{{bai2009penalized}}. It covers many important statistical models, such as the single-index model and the partially linear model. When $\bm\theta_0=\bm0$ or, equivalently, there are no predictors $Z_{ij}$, model (\ref{PLSIM}) is a longitudinal single-index model with an unknown link function. The appeal of the model is that by focusing on an index $X_{ij}^{{\rm T}}\bm\beta_0$, the so-called ``curse of dimensionality" in fitting multivariate nonparametric regression functions is avoided. \cite{chiou2005estimated} introduced a flexible marginal modeling approach and proposed the estimated estimating equations (EEE) method to estimate the index parameter vector $\bm\beta_0$. \cite{lai2012bias} used the smooth threshold GEE method to do variable selection for this model. When $p=1$ and $\beta_0=1$, model (\ref{PLSIM}) becomes the longitudinal partially linear model, which has been investigated in \cite{zeger1994semipar}, \cite{lin2001semiparametric}, \cite{he2002estimation}, \cite{fan2004new}, \cite{sun2003iterative}, \cite{you2007statistical}, \cite{xue2007empirical}, \cite{fan2007analysis}, \cite{li2008generalized} and the references therein. When $m_{i}=1$, model (\ref{PLSIM}) is reduced to the non-longitudinal partially linear single-index model, studies of which include \cite{cui2011efm}, \cite{liang2010estimation}, \cite{wang2010estimation}, \cite{xia2006semi}, \cite{xue2006empirical}, \cite{zhu2006empirical}, \cite{yu2002penalized}, \cite{carroll1997generalized}, among others.

A popular approach for longitudinal/clustered data analysis is by using GEE \citep{liang1986longitudinal}. Variable selection using GEE has been considered in \cite{johnson2008penalized}, \cite{wang2011penalized} and \cite{li2013sgee}. The QIF method, introduced by \cite{qu2000improving}, is a competitor in analyzing longitudinal data. \cite{qu2006quadratic} applied the QIF method to varying coefficient models for longitudinal data. \cite{Bai2008partially,bai2009penalized} applied the QIF method to partially linear models and single-index models with longitudinal data, without considering variable selection. \cite{wang2009consistent} used BIC for consistent variable selection in the context of QIF. Based on the QIF method, \cite{lai2013QIF} studied the estimation and testing issues for the partially linear single-index model with longitudinal data.

Our work differs from the existing works in two major aspects. First, we consider and compare both GEE and QIF in our study while all previous works on single-index models on longitudinal data only consider one of them. It is of significant interest to compare the two approaches in a single study given their similarities. Second, variable selection for single-index models on longitudinal data has not been considered before and we particularly focus on this aspect in our numerical studies, although we need to spend a lot of efforts in explaining GEE and QIF themselves first.

Compared to the work of \cite{li2010empirical}, although our GEE  method is based on the bias correction idea proposed there, the focus of \cite{li2010empirical} is on empirical likelihood method for inferences. Our GEE estimator without penalization is actually the same as the empirical likelihood estimator since the same estimating equations are used in both cases. Note however that the asymptotic properties of the empirical likelihood estimator were not studied before (\cite{li2010empirical} only studied the Wilks' phenomenon for empirical likelihood ratio under the null hypothesis). Compared to the work of \cite{wang2009consistent}, they only considered variable selection for parametric models and our variable selection procedure involving nonparametric components is more challenging and also requires two penalties.

The rest of the paper is organized as follows. In Section \ref{sec2}, we propose the bias-corrected GEE procedure for the partially linear single-index models with longitudinal data and show its asymptotic properties. Section \ref{sec3} reviews a bias-corrected QIF method that has been previously proposed, and discusses the asymptotic properties for the proposed estimator. In Section \ref{sec4}, the variable selection procedure is presented for this model. In Section \ref{sec6}, we present the empirical results from some simulation studies and a real data analysis to illustrate the proposed methods. Finally,  we conclude the paper in Section \ref{sec7} with some discussions. The technical proofs are contained in the supplementary material.

\section{Bias-corrected GEE estimation}\label{sec2}

Assume that the recorded data
$\{(Y_{ij},X_{ij},Z_{ij}),i=1,\ldots,n,j=1,\ldots,m_{i}\}$ are generated from model (\ref{PLSIM}). The identifiability condition $\|\bm\beta_0\|=1$ means that the true value of $\bm\beta_0$ is a boundary point on the unit sphere, which causes some difficulty in estimation. To solve this problem, we use the popular ``delete-one-component" method (see \cite{xue2006empirical,zhu2006empirical}). Let $\bm\beta=(\beta_{1},\ldots,\beta_{p})^{\rm T}$ and let $\bm\beta^{(r)}=(\beta_{1},\ldots,\beta_{r-1},\beta_{r+1},\ldots,\beta_{p})^{\rm T}$ be a $p-1$ dimensional parameter vector after removing the $r$th component $\beta_{r}$. Without loss of generality, we may assume that the true vector $\bm\beta$ has a positive component $\beta_{r}$ (otherwise, consider $-\bm\beta$). Then, we can write 
\begin{eqnarray}\label{beta_J}
\bm\beta=\bm\beta(\bm\beta^{(r)})=(\beta_{1},\ldots,\beta_{r-1},(1-\|\bm\beta^{(r)}\|^{2})^{1/2},\beta_{r+1},\ldots,\beta_{p})^{\rm
T}.
\end{eqnarray}
The true parameter $\bm\beta^{(r)}$ satisfies the constraint $\|\bm\beta^{(r)}\|<1$. Thus, $\bm\beta$ is infinitely differentiable in a neighborhood of the true parameter $\bm\beta^{(r)}$, and the Jacobian matrix is 
$$J_{\bm\beta^{(r)}}=\frac{\partial\bm\beta}{\partial\bm\beta^{(r)}}=(\gamma_{1},\ldots,\gamma_{p})^{\rm T},$$
where $\gamma_{s}(1\leq s\leq p,s\neq r)$ is a $(p-1)$-dimensional unit vector with $s$th component 1, and $\gamma_{r}=-(1-\|\bm\beta^{(r)}\|^{2})^{-1/2}\bm\beta^{(r)}.$

We first introduce the following matrix notations. Let
\begin{eqnarray*}
&&{\bm X}_{i}=(X_{i1},X_{i2},\ldots,X_{im_{i}})^{\rm T},\hskip0.3cm {\bm Z}_{i}=(Z_{i1},Z_{i2},\ldots,Z_{im_{i}})^{\rm T},\\
&&{\bm Y}_{i}=(Y_{i1},Y_{i2},\ldots,Y_{im_{i}})^{\rm T},\hskip0.7cm G({\bm X}_{i}\bm\beta_0)=(g(X_{i1}^{\rm T}\bm\beta_0),g(X_{i2}^{\rm
T}\bm\beta_0),\ldots,g(X_{im_{i}}^{\rm T}\bm\beta_0))^{\rm T}.
\end{eqnarray*}
Let $g_{1}(t)=E[X_{ij}|X_{ij}^{{\rm T}}\bm\beta_0=t]$ and $g_{2}(t)=E[Z_{ij}|X_{ij}^{{\rm T}}\bm\beta_0=t]$. Motivated by the idea of bias correction (\cite{zhu2006empirical,li2010empirical}) and the idea of GEE (\cite{liang1986longitudinal}), we construct the bias-corrected GEE as
\begin{equation}\label{ARV}
\mathcal{Q}_n(G,g_1,g_2,\bm\beta^{(r)},\bm\theta)=:\sum_{i=1}^n\Lambda_{i}V_{i}^{-1}[{\bm
Y}_{i}-G({\bm X}_{i}\bm\beta)-{\bm Z}_{i}\bm\theta]=0,
\end{equation}
where
$$\Lambda_{i}=\left(
                \begin{array}{cc}
                  g'(X_{i1}^{\rm T}\bm\beta)(X_{i1}-g_1(X_{i1}^{\rm T}\bm\beta))^{\rm T}J_{\bm\beta^{(r)}} & (Z_{i1}-g_2(X_{i1}^{\rm T}\bm\beta))^{\rm T} \\
                  g'(X_{i2}^{\rm T}\bm\beta)(X_{i2}-g_1(X_{i2}^{\rm T}\bm\beta))^{\rm T}J_{\bm\beta^{(r)}} & (Z_{i2}-g_2(X_{i2}^{\rm T}\bm\beta))^{\rm T} \\
                  \vdots & \vdots \\
                  g'(X_{im_i}^{\rm T}\bm\beta)(X_{im_i}-g_1(X_{im_i}^{\rm T}\bm\beta))^{\rm T}J_{\bm\beta^{(r)}} & (Z_{im_i}-g_2(X_{im_i}^{\rm T}\bm\beta))^{\rm T} \\
                \end{array}
              \right)^{\rm T}$$
              and $g^{\prime}(\cdot)$ is the
derivative of $g(\cdot)$. For the bias-corrected GEE (\ref{ARV}), $V_{i}={A}_i^{1/2}{R}_i(\alpha){A}_i^{1/2}$ is an invertible working covariance matrix with $A_{i}$ being the $m_i\times m_i$ diagonal matrix of marginal variances and $R_i(\alpha)$ being the working correlation matrix, where $\alpha$ is a vector which fully characterizes $R_i(\alpha)$. Note that $V_i$ will be equal to ${\rm Cov}({\bm Y}_i)$ if ${R}_i$ is indeed the true correlation matrix for ${\bm Y}_i$. Some common working correlation structures include independent structure, compound symmetry (CS) (i.e., exchangeable) with ${R}_{ij}=\rho$ for any $i\neq j$, or first-order autoregressive (AR(1)) with ${R}_{ij}=\rho^{|i-j|}$, where ${R}_{ij}$ denotes the $(i,j)$th element of ${R}$. If the working covariance matrix $V_{i}=I_{m_i}$ is used, with $I_{m_i}$ the $m_i\times m_i$ identity matrix, we ignore the dependence of the data within a cluster, that is,  assume working independence (see \cite{lin2001semiparametric}); when $V_{i}=\Sigma_{i}$, it assumes the true within-subject correlation structure for longitudinal data. In practice, the working covariance matrix $V_{i}$ can be estimated by using the method of moments (\cite{liang1986longitudinal}).

When $g(\cdot)$, $g'(\cdot), g_1(\cdot)$ and $g_2(\cdot)$ are known, we can obtain the estimators of $\bm\beta_0^{(r)}$ and $\bm\theta_0$ by solving the above bias-corrected GEE (\ref{ARV}) directly. However, these quantities in the bias-corrected GEE (\ref{ARV}) are unknown. To obtain the estimators of $\bm\beta_0^{(r)}$ and $\bm\theta_0$, we need to replace them by their estimates.

For given $(\bm\beta,\bm\theta)$, we first apply the local linear smoother \citep{fan1996local} to estimate $g(\cdot)$ and $g'(\cdot)$ by entirely ignoring the within-subject correlation. \cite{lin2000nonparametric} showed that, when standard kernel methods are used, correctly specifying the correlation matrix in fact will result in an asymptotically less efficient estimator for the nonparametric part. We find ($a$, $b$) that minimize
\begin{equation}\label{WLS}
\sum_{i=1}^{n}\sum_{j=1}^{m_{i}}(Y_{ij}-Z_{ij}^{{\rm T}}\bm\theta-a-b(X_{ij}^{{\rm T}}\bm\beta-t))^{2} K_{h}(X_{ij}^{{\rm
T}}\bm\beta-t),
\end{equation}
where $K_{h}(\cdot)=h^{-1}K(\cdot/h),K(\cdot)$ is a kernel function and $h=h_{n}$ is a sequence of positive numbers tending to zero, called the bandwidth. Let $(\hat{a},\hat{b})$ be the solution to the weighted least squares problem (\ref{WLS}). We define the estimators $\hat{g}(t;\bm\beta,\bm\theta)=\hat{a}$ and $\hat{g}^{\prime}(t;\bm\beta,\bm\theta)=\hat{b}$. Simple calculations yield
\begin{equation}\label{GE}
\hat{g}(t;\bm\beta,\bm\theta)=\sum_{i=1}^{n}\sum_{j=1}^{m_{i}}W_{nij}(t;\bm\beta)(Y_{ij}-Z_{ij}^{\rm
T}\bm\theta),
\end{equation}
and
\begin{equation}\label{GDE}
\hat{g}^{\prime}(t;\bm\beta,\bm\theta)=\sum_{i=1}^{n}\sum_{j=1}^{m_{i}}\widetilde{W}_{nij}(t;\bm\beta)(Y_{ij}-Z_{ij}^{\rm
T}\bm\theta),
\end{equation}
where
\begin{equation*}\label{26}
W_{nij}(t;\bm\beta)=\frac{N^{-1}K_{h}(X_{ij}^{\rm
T}\bm\beta-t)[S_{n,2}(t;\bm\beta)-(X_{ij}^{\rm
T}\bm\beta-t)S_{n,1}(t;\bm\beta)]}
{S_{n,0}(t;\bm\beta)S_{n,2}(t;\bm\beta)-S_{n,1}^{2}(t;\bm\beta)},
\end{equation*}
\begin{equation*}\label{27}
\widetilde{W}_{nij}(t;\bm\beta)=\frac{N^{-1}K_{h}(X_{ij}^{\rm
T}\bm\beta-t)[(X_{ij}^{\rm
T}\bm\beta-t)S_{n,0}(t;\bm\beta)-S_{n,1}(t;\bm\beta)]}
{S_{n,0}(t;\bm\beta)S_{n,2}(t;\bm\beta)-S_{n,1}^{2}(t;\bm\beta)},
\end{equation*}
 and
$$S_{n,l}(t;\bm\beta)=\frac{1}{N}\sum_{i=1}^{n}\sum_{j=1}^{m_{i}}K_{h}(X_{ij}^{\rm T}\bm\beta-t)(X_{ij}^{\rm T}\bm\beta-t)^{l},\quad l=0,1,2.$$

Given $\bm\beta$, we can obtain the estimators of $g_{1}(\cdot)$ and $g_{2}(\cdot)$ as 
\begin{equation}\label{G1}
\hat{g}_{1}(t;\bm\beta)=\sum_{i=1}^{n}\sum_{j=1}^{m_{i}}W_{nij}(t;\bm\beta)X_{ij},
\end{equation}
and
\begin{equation}\label{G2}
\hat{g}_{2}(t;\bm\beta)=\sum_{i=1}^{n}\sum_{j=1}^{m_{i}}W_{nij}(t;\bm\beta)Z_{ij}.
\end{equation}

We estimate the $(p-1+q)$-dimensional parameter vector $\bm\xi_0=(\bm\beta_0^{(r){\rm T}},\bm\theta_0^{\rm T})^{\rm T}$ by solving the following estimated bias-corrected GEE 
\begin{equation}\label{BCARV}
\mathcal{Q}_n(\hat{G},\hat{g}_1,\hat{g}_2,\bm\beta^{(r)},\bm\theta)=\sum_{i=1}^n\hat{\Lambda}_{i}V_{i}^{-1}[{\bm
Y}_{i}-\hat{G}({\bm X}_{i}{\bm\beta};\bm\beta,\bm\theta)-{\bm
Z}_{i}{\bm\theta}]=0,
\end{equation}
 where $\hat{G}({\bm X}_{i}{\bm\beta};\bm\beta,\bm\theta)=(\hat{g}(X_{i1}^{{\rm T}}{\bm\beta};\bm\beta,\bm\theta),\hat{g}(X_{i2}^{{\rm
T}}{\bm\beta};\bm\beta,\bm\theta), \ldots,\hat{g}(X_{im_{i}}^{{\rm T}}{\bm\beta};\bm\beta,\bm\theta))^{{\rm T}}$ and 
$$\hat{\Lambda}_{i}=\left(
                \begin{array}{cc}
                  \hat{g}^{\prime}(X_{i1}^{{\rm
T}}\bm\beta;\bm\beta,\bm\theta)(X_{i1}-\hat{g}_{1}(X_{i1}^{{\rm
T}}\bm\beta;\bm\beta))^{{\rm T}}J_{\bm\beta^{(r)}} & (Z_{i1}-\hat{g}_2(X_{i1}^{\rm T}\bm\beta;\bm\beta))^{\rm T} \\
                  \hat{g}^{\prime}(X_{i2}^{{\rm
T}}\bm\beta;\bm\beta,\bm\theta)(X_{i2}-\hat{g}_{1}(X_{i2}^{{\rm
T}}\bm\beta;\bm\beta))^{{\rm T}}J_{\bm\beta^{(r)}} & (Z_{i2}-\hat{g}_2(X_{i2}^{\rm T}\bm\beta;\bm\beta))^{\rm T} \\
                  \vdots & \vdots \\
                  \hat{g}'(X_{im_i}^{\rm T}\bm\beta;\bm\beta,\bm\theta)(X_{im_i}-\hat{g}_1(X_{im_i}^{\rm T}\bm\beta;\bm\beta))^{\rm T}J_{\bm\beta^{(r)}} & (Z_{im_i}-\hat{g}_2(X_{im_i}^{\rm T}\bm\beta;\bm\beta))^{\rm T} \\
                \end{array}
              \right)^{\rm T}.$$

The Newton-Raphson iterative algorithm can be used to solve the bias-corrected GEE (\ref{BCARV}) and find the estimators of $\bm\beta_0^{(r)}$ and $\bm\theta_0$. The iterative algorithm is described as follows.

{\it Step 1}: Start with initial estimators of $\bm\beta_0$ and $\bm\theta_0$, say $\hat{\bm\beta}_{(0)}$ and $\hat{\bm\theta}_{(0)}$.

{\it Step 2}: Use the current estimates $\hat{\bm\beta}_{(k)}$ and $\hat{\bm\theta}_{(k)}$ and (\ref{GE})--(\ref{G2}) to obtain the estimators $\hat{g}(t;\hat{\bm\beta}_{(k)},\hat{\bm\theta}_{(k)}), \hat{g}^{\prime}(t;\hat{\bm\beta}_{(k)},\hat{\bm\theta}_{(k)}), \hat{g}_{1}(t;\hat{\bm\beta}_{(k)})$ and $\hat{g}_{2}(t;\hat{\bm\beta}_{(k)})$. Based on these estimators, compute the working covariance matrix $\hat{V}_{i, (k)}$.

{\it Step 3}: Use these estimates of $g(\cdot)$, $g'(\cdot), g_1(\cdot)$,  $g_2(\cdot)$ and $V_i$ from Step 2 and the estimated bias-corrected GEE (\ref{BCARV}) to obtain the updated estimate $\hat{\bm\xi}_{(k+1)}=(\hat{\bm\beta}_{(k+1)}^{(r){\rm T}},\hat{\bm\theta}_{(k+1)}^{\rm T})^{\rm T}$. Compute 
\begin{eqnarray}\label{fisher-est}
\hat{\bm\xi}_{(k+1)}=\hat{\bm\xi}_{(k)}+{\Pi}_n^{-1}(\hat{\bm\beta}_{(k)}^{(r)},\hat{\bm\theta}_{(k)})\mathcal{Q}_n(\hat{G},\hat{g}_1,\hat{g}_2,\hat{\bm\beta}_{(k)}^{(r)},\hat{\bm\theta}_{(k)}),
\end{eqnarray}
$$\Pi_n(\hat{\bm\beta}_{(k)}^{(r)},\hat{\bm\theta}_{(k)})=\sum_{i=1}^{n}\hat{\Lambda}_i(\hat{\bm\beta}_{(k)}^{(r)},\hat{\bm\theta}_{(k)})\hat{V}_{i,(k)}^{-1}\hat{\Lambda}_i^{\rm T}(\hat{\bm\beta}_{(k)}^{(r)},\hat{\bm\theta}_{(k)}).$$ 
By (\ref{beta_J}) and $\hat{\bm\beta}_{(k+1)}^{(r)}$, obtain the updated estimate $\hat{\bm\beta}_{(k+1)}=\bm\beta(\hat{\bm\beta}_{(k+1)}^{(r)})$.

{\it Step 4}: Repeat the above two steps until the successive value satisfies $\|\hat{\bm\xi}_{(k+1)}-\hat{\bm\xi}_{(k)}\|<\epsilon,$ where $\epsilon$ is some given tolerance value. Denote the final estimator of $\bm\xi_0$ as the bias-corrected GEE estimator.

It is noteworthy that we apply the Newton-Raphson iterative method to find the final estimator $\hat{\bm\xi}=(\hat{\bm\beta}^{(r){\rm T}},\hat{\bm\theta}^{\rm T})^{\rm T}$ of ${\bm\xi}_0=({\bm\beta}_0^{(r){\rm T}},{\bm\theta}_0^{\rm T})^{\rm T}$. Further, our final estimator for $\bm\xi^*=(\bm\beta^{\rm T},\bm\theta^{\rm T})^{\rm T}$ is $\hat{\bm\xi}^*=(\hat{\bm\beta}^{\rm T},\hat{\bm\theta}^{\rm T})^{\rm T}=(\bm\beta(\hat{\bm\beta}^{(r)})^{\rm T},\hat{\bm\theta}^{\rm T})^{\rm T}$. We also define the estimator of the link function $g(\cdot)$ by (\ref{GE}) with $\bm\beta$ and $\bm\theta$ being replaced by $\hat{\bm\beta}$ and $\hat{\bm\theta}$, respectively.

{\remark\label{rem1} In Step 1, consistent initial estimators of $\bm\beta_0$ and $\bm\theta_0$ are needed to help us obtain the final root-$n$ consistent estimators $\hat{\bm\beta}$ and $\hat{\bm\theta}$. The PLSIM algorithm proposed in \cite{xia2006semi} is applied to obtain the initial estimators of $\bm\beta_0$ and $\bm\theta_0$ by ignoring the within-subject correlation. In Section \ref{sec6}, our simulation study shows that the initial estimators perform well. }

In order to study the asymptotic behavior of the proposed estimators, we first give a set of conditions for the results stated in the theorems.
\par
\indent{\bf C1}. \ For any $i=1,\ldots,n,j=1,\ldots,m_{i},$ the density function of $X_{ij}^{\rm T}\bm\beta$ is bounded away from zero and infinity on $\mathcal{T}$ and satisfies the Lipschitz condition of order 1 on $\mathcal{T}$, where $\mathcal{T}=\{t=X_{ij}^{\rm T}\bm\beta: X_{ij}\in A,i=1,\ldots,n,j=1,\ldots,m_{i}\}$ and $ A$ is the compact support set of $X_{ij}$.

\par
\indent{\bf C2}. \ $g(t)$ has two bounded and continuous derivatives on $\mathcal{T}$; $g_{1s}(t)$ and $g_{2k}(t)$ satisfy the local Lipschitz condition of order 1, where $g_{1s}(t)$ and $g_{2k}(t)$ are the $s$th and $k$th component of $g_{1}(t)$ and $g_{2}(t)(1\leq s\leq p,1\leq k\leq q)$ respectively.

\par
\indent{\bf C3}. \ The kernel $K(u)$ is a bounded and symmetric probability density function and satisfies 
$$\int_{-\infty}^{\infty}u^{2}K(u)du\neq0,\quad \int_{-\infty}^{\infty}|u|^{i}K(u)du<\infty,\quad i=1,2,\ldots.$$

\par
\indent{\bf C4}. \ There exists a positive constant $M$ such that $\displaystyle\max_{1\leq i\leq n,1\leq j\leq m_{i}}\sup_{x,z}E(e_{ij}^{4}|X_{ij}=x,Z_{ij}=z)\leq M<\infty$ and $\displaystyle\max_{1\leq i\leq n,1\leq j\leq m_{i}}\sup_{x}E(e_{ij}^{4}|X_{ij}=x)\leq M<\infty$.

\indent{\bf C5}. \ When $n\rightarrow\infty$, the bandwidth $h$ satisfies that $h\rightarrow0,~n^2h^{7}\rightarrow\infty,~nh^{8}\rightarrow0.$

\indent{\bf C6}. \ There exist two positive constants $c_{1}$ and $c_{2}$ such that
$$
0<c_{1}\leq\min_{1\leq i\leq n}\lambda_{i1}\leq\max_{1\leq i\leq n}\lambda_{im_{i}}\leq c_{2}<\infty,
$$
where $\lambda_{i1}$ and $\lambda_{im_{i}}$ denote the smallest and largest eigenvalue of $\Sigma_{i}$, respectively.

\indent{\bf C7}. \ There exist positive constants $c_{3}$ and $c_{4}$ such that
$$
0<c_{3}\leq\min_{1\leq i\leq n}\lambda_{i1}^{\prime}\leq\max_{1\leq i\leq n}\lambda_{im_{i}}^{\prime}\leq c_{4}<\infty,
$$
where $\lambda_{i1}^{\prime}$ and $\lambda_{im_{i}}^{\prime}$ denote the smallest and largest eigenvalue of $V_{i}$, respectively.

\indent{\bf C8}. \ There exists a positive constant $M$ such that for all $i,j$, $\sup\limits_{t\in \mathcal{T}}E(\|Z_{ij}\|^{2}|X_{ij}^{\rm T}\bm\beta=t)\leq M<\infty$.

\indent{\bf C9}. \
$\Omega(\bm\beta_0^{(r)},\bm\theta_0)=\displaystyle\lim_{n\rightarrow\infty}\frac{1}{n}\sum_{i=1}^{n}E\{\Lambda_{i}V_{i}^{-1}({\bm e}_{i}{\bm e}_{i}^{\rm T})V_{i}^{-1} \Lambda_{i}^{\rm T}\}$ and $\Pi=\displaystyle\lim_{n\rightarrow\infty}\frac{1}{n}\sum_{i=1}^{n}E\Big[\Lambda_iV_i^{-1}\Lambda_i^{\rm T}\Big]$ are two positive definite matrices, where $\Lambda_{i}$  is defined in (\ref{ARV}).

\indent{\bf C10}. \ The matrix $\frac{1}{n}\sum_{i=1}^{n}U_i(\bm\beta_0^{(r)},\bm\theta_0)U_i^{\rm T}(\bm\beta_0^{(r)},\bm\theta_0)$ converges almost surely to an invertible positive definite matrix $\Sigma(\bm\beta_0^{(r)},\bm\theta_0)$, where
\begin{small}
\begin{eqnarray*}
{U}_i(\bm\beta_0^{(r)},\bm\theta_0)=\left(
                                                                              \begin{array}{c}
                                                                                {\Lambda}_{i}{A}_{i}^{-1/2}M_1{A}_{i}^{-1/2}{\bm e}_{i} \\
                                                                                {\Lambda}_{i}{A}_{i}^{-1/2}M_2{A}_{i}^{-1/2}{\bm e}_{i} \\
                                                                                \vdots \\
                                                                                {\Lambda}_{i}{A}_{i}^{-1/2}M_k{A}_{i}^{-1/2}{\bm e}_{i} \\
                                                                              \end{array}
                                                                            \right),\quad
\Sigma(\bm\beta_0^{(r)},\bm\theta_0)=\lim_{n\rightarrow\infty}\frac{1}{n}\sum_{i=1}^{n}E[{U}_i(\bm\beta_0^{(r)},\bm\theta_0){U}_i^{\rm
T}(\bm\beta_0^{(r)},\bm\theta_0)].
\end{eqnarray*}
\end{small}

Conditions C1--C8 are actually quite mild and can be easily satisfied, and these conditions are also found in \cite{li2010empirical}.  Condition  C1 ensures that the denominators of $\hat{g}(t;\bm\beta,\bm\theta)$ and $\hat{g}^{\prime}(t;\bm\beta,\bm\theta)$ are, with high probability, bounded away from 0 on $t\in \mathcal{T}$ for $\bm\beta$ in a neighborhood of $\bm\beta_0$. Condition C2 is the standard smoothness condition. Condition C3 is the usual assumption for the kernel function. Condition C4 is a necessary condition for the consistency and the asymptotic normality of the estimator.  Condition C5 allows a range of bandwidths that include the optimal bandwidth because the bias-corrected technique is used. Therefore, we do not need to use different bandwidths to estimate $g(\cdot)$ and $g^{\prime}(\cdot)$. Conditions C6 and C7 ensure that the covariance matrix $\Sigma_i$ and the working covariance matrix $V_i$ are invertible for $i=1,\ldots,n$. Condition C8 is a technical condition on the moments of the predictors. Conditions C9 and C10 ensure that the asymptotic variances exist for the bias-corrected GEE estimator and the bias-corrected QIF estimator, respectively.

{\theo\label{theo1} Suppose that the technical conditions (C1)--(C9) hold, and the $r$th component of $\bm\beta_0$ is positive. Further suppose that the initial estimator is $\sqrt{n}$-consistent (initial estimator can be obtained, for example, as in \cite{xia2006semi}), then there exist solutions $\hat{\bm\beta}^{(r)}$ and $\hat{\bm\theta}$ of (\ref{BCARV}) that satisfy
 \begin{eqnarray*}
 \sqrt{n}\left(
                                                      \begin{array}{c}
                                                        \hat{\bm\beta}^{(r)}-{\bm\beta}_0^{(r)} \\
                                                        \hat{\bm\theta}-\bm\theta_0 \\
                                                      \end{array}\right)\stackrel{L}\longrightarrow
                                                      N(0,\Pi^{-1}\Omega(\bm\beta_0^{(r)},\bm\theta_0)\Pi^{-1}),
 \end{eqnarray*}
where ``$\stackrel{L}\longrightarrow$" stands for convergence in distribution, and $\Omega(\bm\beta_0^{(r)},\bm\theta_0)$ and $\Pi$ are the positive matrices defined in condition C9.}

We now consider the asymptotic normality of the estimator $(\hat{\bm\beta}^{\rm T},\hat{\bm\theta}^{\rm T})^{\rm T}$. By the result of \cite{wang2010estimation}, we have
\begin{eqnarray*}
\hat{\bm\beta}-\bm\beta_0=J_{\bm\beta_0^{(r)}}(\hat{\bm\beta}^{(r)}-\bm\beta_0^{(r)})+O_P(n^{-1}).
\end{eqnarray*}
By Theorem \ref{theo1} and the  Slutsky's Theorem, we have the following result.

{\coll\label{coll1} Under the conditions of Theorem \ref{theo1}, we have
 \begin{eqnarray*}
 \sqrt{n}\left(
                                                      \begin{array}{c}
                                                        \hat{\bm\beta}-{\bm\beta}_0 \\
                                                        \hat{\bm\theta}-\bm\theta_0 \\
                                                      \end{array}\right)\stackrel{L}\longrightarrow
                                                      N(0,\Psi),
 \end{eqnarray*}
 where $$\Psi=\left(
                \begin{array}{cc}
                  J_{\bm\beta_0^{(r)}} & {\bm 0}_{p\times q} \\
                  {\bm 0}_{q\times(p-1)} & I_{q} \\
                \end{array}
              \right)
 \Pi^{-1}\Omega(\bm\beta_0^{(r)},\bm\theta_0)\Pi^{-1}\left(
                \begin{array}{cc}
                  J_{\bm\beta_0^{(r)}} & {\bm 0}_{p\times q} \\
                  {\bm 0}_{q\times(p-1)} & I_{q} \\
                \end{array}
              \right)^{\rm T}.$$}

\section{Bias-corrected QIF}\label{sec3}

As the working covariance matrix $V_i$ is unknown in practice, misspecification of the working covariance matrix $V_i$ will lead to less efficient estimators of regression coefficients. To improve the efficiency of estimation, \cite{qu2000improving} introduced the QIF by assuming that the inverse of the working correlation can be approximated by a linear combination of several basis matrices, that is 
\begin{equation}\label{qif}
{R}^{-1}\approx a_1M_1+a_2M_2+\cdots+a_kM_k,
\end{equation}
where $M_1$ is the identity matrix, and $M_2,\ldots,M_k$ are symmetric basis matrices which are determined by the structure of ${R}$, and $a_1,\ldots,a_k$ are constant coefficients. The advantage of this approach is that it does not require estimation of linear coefficients $a_i$'s which can be viewed as nuisance parameters. In practice, we need to choose the basis matrices $M_1,\ldots,M_k$. If the correlation matrix ${R}$ is exchangeable, then ${R}^{-1}=a_1M_1+a_2M_2$, where $M_1$ is the identity matrix and $M_2$ is a matrix with 0 on the diagonal and 1 off-diagonal. If the correlation matrix ${R}$ is AR(1), then ${R}^{-1}=a_1M_1^*+a_2M_2^*+a_3M_3^*$, where $M_1^*$ is the identity matrix, $M_2^*$ has 1 on the sub-diagonal and 0 elsewhere, and $M_3^*$ has 1 on the corners (1,1) and $(m,m)$ and 0 elsewhere (\cite{qu2000improving}, \cite{qu2006quadratic}). \cite{qu2003building} developed an adaptive estimating equation approach to find a reliable approximation to the inverse of the variance matrix.

Based on the bias-corrected GEE (\ref{BCARV}) and (\ref{qif}), \cite{lai2013QIF} defined the following bias-corrected QIF objective function
\begin{eqnarray}\label{BC-QIF}
{Q}_n(\bm\beta^{(r)},\bm\theta)=\overline{\hat{U}}_n^{\rm T}(\bm\beta^{(r)},\bm\theta)C_n^{-1}(\bm\beta^{(r)},\bm\theta)\overline{\hat{U}}_n(\bm\beta^{(r)},\bm\theta),
\end{eqnarray}
where $\overline{\hat{U}}_n(\bm\beta^{(r)},\bm\theta)=n^{-1}\sum\limits_{i=1}^n\hat{U}_i(\bm\beta^{(r)},\bm\theta)$, $C_n(\bm\beta^{(r)},\bm\theta)=n^{-1}\sum\limits_{i=1}^n\hat{U}_i(\bm\beta^{(r)},\bm\theta)\hat{U}_i^{\rm T}(\bm\beta^{(r)},\bm\theta)$ and
\begin{eqnarray}\label{equ-qif}
\hat{U}_i(\bm\beta^{(r)},\bm\theta)=\left(
                                                                              \begin{array}{c}
                                                                                \hat{\Lambda}_{i}{A}_{i}^{-1/2}M_1{A}_{i}^{-1/2}[{\bm Y}_{i}-\hat{G}({\bm
X}_{i}\bm\beta;\bm\beta,\bm\theta)-{\bm Z}_{i}\bm\theta] \\
                                                                                \hat{\Lambda}_{i}{A}_{i}^{-1/2}M_2{A}_{i}^{-1/2}[{\bm Y}_{i}-\hat{G}({\bm
X}_{i}\bm\beta;\bm\beta,\bm\theta)-{\bm Z}_{i}\bm\theta] \\
                                                                                \vdots \\
                                                                                \hat{\Lambda}_{i}{A}_{i}^{-1/2}M_k{A}_{i}^{-1/2}[{\bm Y}_{i}-\hat{G}({\bm
X}_{i}\bm\beta;\bm\beta,\bm\theta)-{\bm Z}_{i}\bm\theta] \\
                                                                              \end{array}
                                                                            \right).
\end{eqnarray}
It is easy to check that the bias-corrected GEE defined in (\ref{BCARV}) becomes a linear combination of the extended score vector $\sum_{i=1}^{n}\hat{U}_i(\bm\beta^{(r)},\bm\theta)$. Note that the dimension of $\hat{U}_i(\bm\beta^{(r)},\bm\theta)$ is $l=k(p-1+q)$, and it is greater than the number of unknown parameters.  Thus,   the method of GMM proposed by \cite{hansen1982large} can be extended to obtain the estimators of $\bm\beta_0^{(r)}$ and $\bm\theta_0$ by minimizing the bias-corrected QIF (\ref{BC-QIF}). Similarly, the Newton-Raphson iterative algorithm can be also used to  find the estimators of $\bm\beta_0^{(r)}$ and $\bm\theta_0$.

%
%
%
%
%

Let $(\hat{\bm\beta}_*^{(r){\rm T}},\hat{\bm\theta}_*^{\rm T})^{\rm T}$ be the bias-corrected QIF estimator of $\bm\xi_0=(\bm\beta_0^{(r){\rm T}},\bm\theta_0^{\rm T})^{\rm T}$, then our  bias-corrected QIF estimator for $(\bm\beta^{\rm T},\bm\theta^{\rm T})^{\rm T}$ is $(\hat{\bm\beta}_*^{\rm T},\hat{\bm\theta}_*^{\rm T})^{\rm T}=(\bm\beta(\hat{\bm\beta}_*^{(r)})^{\rm T},\hat{\bm\theta}_*^{\rm T})^{\rm T}$. We also define the estimator of the link function $g(\cdot)$ by (\ref{GE}) with $\bm\beta$ and $\bm\theta$ replaced by $\hat{\bm\beta}_*$ and $\hat{\bm\theta}_*$, respectively. The following asymptotic results have been obtained in \cite{lai2013QIF}.

{\theo\label{theo2} Suppose that the technical conditions (C1)--(C8) and (C10) hold, then we have

{\rm (1)}\ \  the bias-corrected QIF estimator $(\hat{\bm\beta}_*^{(r){\rm T}},\hat{\bm\theta}_*^{\rm T})^{\rm T}$ by minimizing (\ref{BC-QIF}) exists and converges to $({\bm\beta}_0^{(r){\rm T}},{\bm\theta}_0^{\rm T})^{\rm T}$ in probability;

{\rm (2)}\ \ the bias-corrected QIF estimator $(\hat{\bm\beta}_*^{(r){\rm T}},\hat{\bm\theta}_*^{\rm T})^{\rm T}$ is asymptotically normal. That is
 \begin{eqnarray*}
 \sqrt{n}\left(
                                                      \begin{array}{c}
                                                        \hat{\bm\beta}_*^{(r)}-{\bm\beta}_0^{(r)} \\
                                                        \hat{\bm\theta}_*-\bm\theta_0 \\
                                                      \end{array}\right)\stackrel{L}\longrightarrow
                                                      N\left(0,\Big(\Gamma^{\rm
T}\Sigma^{-1}(\bm\beta_0^{(r)},\bm\theta_0)\Gamma\Big)^{-1}\right),
\end{eqnarray*}
 where  $\Sigma(\bm\beta_0^{(r)},\bm\theta_0)$ is
the positive definite matrix defined in conditions (C10), and
$$\Gamma=\lim_{n\rightarrow\infty}\frac{1}{n}\sum_{i=1}^{n}E\left(
                                                                              \begin{array}{c}
                                                                                {\Lambda}_{i}{A}_{i}^{-1/2}M_1{A}_{i}^{-1/2}{\Lambda}_{i}^{\rm T} \\
                                                                                {\Lambda}_{i}{A}_{i}^{-1/2}M_2{A}_{i}^{-1/2}{\Lambda}_{i}^{\rm T} \\
                                                                                \vdots \\
                                                                                {\Lambda}_{i}{A}_{i}^{-1/2}M_k{A}_{i}^{-1/2}{\Lambda}_{i}^{\rm T} \\
                                                                              \end{array}
                                                                            \right)$$ is an $l\times (p-1+q)$ matrix with the rank being
$p-1+q$.}

By $\hat{\bm\beta}_*-\bm\beta_0=J_{\bm\beta_0^{(r)}}(\hat{\bm\beta}_*^{(r)}-\bm\beta_0^{(r)})+O_P(n^{-1})$, we have the following asymptotic normality of the estimator $(\hat{\bm\beta}_*^{\rm T},\hat{\bm\theta}_*^{\rm T})^{\rm T}$.

{\coll\label{coll2} Under the conditions of Theorem \ref{theo2}, we have
 \begin{eqnarray*}
 \sqrt{n}\left(
                                                      \begin{array}{c}
                                                        \hat{\bm\beta}_*-{\bm\beta}_0 \\
                                                        \hat{\bm\theta}_*-\bm\theta_0 \\
                                                      \end{array}\right)\stackrel{L}\longrightarrow
                                                      N(0,\Phi),
 \end{eqnarray*}
 where $$\Phi=\left(
                \begin{array}{cc}
                  J_{\bm\beta_0^{(r)}} & {\bm 0}_{p\times q} \\
                  {\bm 0}_{q\times(p-1)} & I_{q} \\
                \end{array}
              \right)
 \Big(\Gamma^{\rm
T}\Sigma^{-1}(\bm\beta_0^{(r)},\bm\theta_0)\Gamma\Big)^{-1}\left(
                \begin{array}{cc}
                  J_{\bm\beta_0^{(r)}} & {\bm 0}_{p\times q} \\
                  {\bm 0}_{q\times(p-1)} & I_{q} \\
                \end{array}
              \right)^{\rm T}.$$}

\section{Variable selection and the asymptotic properties}\label{sec4}

In practice, not all explanatory variables are predictive of the response. It is of interest to automatically select the relevant predictors in the model. We use penalization approach to simultaneously estimate parameters and remove irrelevant variables. Given $q_\lambda=p_\lambda'$ for some penalty function $p_\lambda$, we consider the bias-corrected penalized GEE
\begin{small}
\begin{equation}\label{eqn:pengee}
U^P(\bm\beta^{(r)},\bm\theta)=\sum_{i=1}^n\hat{\Lambda}_{i}V_{i}^{-1}[{\bm Y}_{i}-\hat{G}({\bm X}_{i}\bm\beta;\bm\beta,\bm\theta)-{\bm Z}_{i}\bm\theta]-n\bm{q}_{\lambda_1}(|\bm\beta^{(r)}|) {\bf sgn}(\bm\beta^{(r)})-n\bm{q}_{\lambda_2}(|\bm\theta|){\bf sgn}(\bm\theta),
\end{equation}
\end{small}
where
$$\bm{q}_{\lambda_1}(|\bm\beta^{(r)}|)=(q_{\lambda_1}(|\beta_1|),\ldots,q_{\lambda_1}(|\beta_{r-1}|),q_{\lambda_1}(|\beta_{r+1}|), \ldots,q_{\lambda_1}(|\beta_{p}|),\bm{0}_{q\times 1}^{\rm T})^{\rm T},$$
$${\bf{sgn}}(\bm\beta^{(r)})=({\rm{sgn}}(\beta_1),\ldots,{\rm sgn}(\beta_{r-1}),{\rm{sgn}}(\beta_{r+1}),\ldots,{\rm{sgn}}(\beta_p),{\bf0}_{q\times 1}^{\rm T})^{\rm T},$$ 
with $\mbox{sgn}(t)=I(t>0)-I(t<0)$ the sign function, and $\bm{q}_\lambda(|\bm\beta^{(r)}|){\bf{sgn}}(\bm\beta^{(r)})$ is the
componentwise product of $\bm{q}_\lambda(|\bm\beta^{(r)}|)$ and ${\bf{sgn}}(\bm\beta^{(r)})$. Similarly, $\bm{q}_{\lambda_2}(|\bm\theta|)=({\bf{0}}_{(p-1)\times 1}^{\rm T},q_{\lambda_2}(|\theta_1|),\ldots,q_{\lambda_2}(|\theta_{q}|))^{\rm T}$, ${\bf{sgn}}(\bm\theta)=({\bm{0}}_{(p-1)\times 1}^{\rm T},{\rm{sgn}}(\theta_1),\ldots,{\rm{sgn}}(\theta_{q}))^{\rm T}$.

Since the penalty is typically not continuous, we consider an approximate zero-crossing of  $U^P(\bm\beta^{(r)},\bm\theta)$. For convenience, we denote $\bm{\xi}_0=({\bm\beta}_0^{(r){\rm T}}, {\bm\theta}_0^{\rm T})^{\rm T}$, $\hat{\bm\xi}=(\hat{\bm\beta}^{(r){\rm T}},\hat{\bm\theta}^{\rm T})^{\rm T}.$ As defined in \cite{johnson2008penalized}, $(\hat{\bm\beta}^{(r)},\hat{\bm\theta})$ is an approximate zero-crossing to (\ref{eqn:pengee}) if $\overline{\lim}_{n\rightarrow \infty}\overline{\lim}_{\epsilon\rightarrow 0+}n^{-1}U^P_j (\hat{\bm\xi}+\epsilon\bm{e}_j)U^P_j(\hat{\bm\xi}-\epsilon\bm{e}_j)\le 0$, $1\le j\le p-1+q$, where $\bm{e}_j$ is the vector with  one at position $j$ and zero otherwise, and $U^P_j$ is the $j$th component of $U^P$.

Various penalty functions have been used in the variable selection literature for linear regression models. We adopt the smoothly clipped absolute deviation (SCAD) penalty function proposed in \cite{fan2001variable}, which is given by
\begin{equation*}\label{scad}
p_{\lambda}^{\prime}(|x|)=\lambda\left\{I(|x|\leq\lambda)+\frac{(c\lambda-|x|)_{+}}{(c-1)\lambda}I(|x|>\lambda)\right\}
\quad  {\rm for~~ some}\quad c>2,
\end{equation*}
where the notation $(z)_{+}$ stands for the positive part of $z$. \cite{fan2001variable} suggested using $c=3.7$ for the SCAD penalty function.

Similarly, for the QIF approach, we can consider the bias-corrected penalized QIF
\begin{equation}\label{eqn:penqif}
 nQ_n(\bm\beta^{(r)},\bm\theta)+n\sum_{1\le j\le p, j\neq r} p_{\lambda_1}(|\beta_j|)+n\sum_{1\le j\le q}p_{\lambda_2}(|\theta_j|).
\end{equation}

To state the theoretical properties of penalized estimators, we assume the parameters in the true model are $\bm\beta_0=(\bm\beta_{1}^{\rm T},\bm\beta_{2}^{\rm T}=\bm{0}^{\rm T})^{\rm T}$ and $\bm\theta_0=(\bm\theta_{1}^{\rm T},\bm\theta_{2}^{\rm T}=\bm{0}^{\rm T})^{\rm T}$, where $\bm\beta_1$ and $\bm\theta_1$ are $p_0$-dimensional and $q_0$-dimensional respectively. We also assume $r\le p_0$, that is, we can correctly identify one nonzero coefficient in the index vector to carry out the ``delete-one-component" procedure.

{\theo\label{theovs}

\noindent (a) Under the conditions (C1)--(C9), if $\lambda_1\rightarrow 0$, $\sqrt{n}\lambda_1\rightarrow \infty$, $\lambda_2\rightarrow 0$, $\sqrt{n}\lambda_2\rightarrow \infty$, then there exists an approximate zero-crossing of the bias-corrected penalized GEE (\ref{eqn:pengee}), denoted by $(\hat{\bm\beta}^{(r)},\hat{\bm\theta})$, that satisfies
\begin{itemize}
\item[(i)] $\hat{\bm\beta}^{(r)}_2=(\hat{\beta}_{p_0+1},\ldots,\hat{\beta}_p)^{\rm T}=\bm{0},\quad \hat{\bm\theta}_2=(\hat{\theta}_{q_0+1},\ldots,\hat{\theta}_q)^{\rm T}=\bm{0}$;

\item[(ii)]  \begin{eqnarray*}
 \sqrt{n}\left(
                                                      \begin{array}{c}
                                                        \hat{\bm\beta}^{(r)}_1-{\bm\beta}^{(r)}_1 \\
                                                        \hat{\bm\theta}_1-\bm\theta_1 \\
                                                      \end{array}\right)\stackrel{L}\longrightarrow
                                                      N(0,\Pi_{11}^{-1}\Omega_{11}(\bm\beta^{(r)}_1,\bm\theta_1)\Pi_{11}^{-1}),
 \end{eqnarray*}
where $\Pi_{11}$, $\Omega_{11}$ are defined similarly as $\Pi$ and $\Omega$, but using only the first $p_0$ columns of ${\bm X}_i$ and the first $q_0$ columns of ${\bm Z}_i$.
\end{itemize}

\noindent (b) Under the conditions (C1)--(C8) and (C10), if $\lambda_1\rightarrow 0$, $\sqrt{n}\lambda_1\rightarrow \infty$, $\lambda_2\rightarrow 0$, $\sqrt{n}\lambda_2\rightarrow \infty$, then there exists a local minimizer of the bias-corrected penalized QIF (\ref{eqn:penqif}), denoted by
$(\hat{\bm\beta}^{(r)},\hat{\bm\theta})$ (with abuse of notation) that satisfies
\begin{itemize}
\item[(i)] $\hat{\bm\beta}^{(r)}_2=(\hat{\beta}_{p_0+1},\ldots,\hat{\beta}_p)^{\rm T}=\bm{0},\quad \hat{\bm\theta}_2=(\hat{\theta}_{q_0+1},\ldots,\hat{\theta}_q)^{\rm T}=\bm{0}$;

\item[(ii)]  \begin{eqnarray*}
 \sqrt{n}\left(
                                                      \begin{array}{c}
                                                        \hat{\bm\beta}_1^{(r)}-{\bm\beta}^{(r)}_1 \\
                                                        \hat{\bm\theta}_1-\bm\theta_1 \\
                                                      \end{array}\right)\stackrel{L}\longrightarrow
                                                      N\left(0,\Big(\Gamma_{11}^{\rm
T}\Sigma_{11}^{-1}(\bm\beta_{1}^{(r)},\bm\theta_{1})\Gamma_{11}\Big)^{-1}\right),
\end{eqnarray*}
where $\Gamma_{11}$, $\Sigma_{11}$ are defined similarly as $\Gamma$ and $\Sigma$, but using only the first $p_0$ columns of ${\bm X}_i$ and the first $q_0$ columns of ${\bm Z}_i$.
\end{itemize}
}

In the process of variable selection, the tuning parameters $\lambda_1$ and $\lambda_2$ should be determined. For a given data set with a finite sample size, it is practically important to select the unknown tuning parameters with a data driven method. In this paper, we use the BIC (\cite{liang2010estimation}) to select the tuning parameters $(\lambda_1,\lambda_2)$, that is 
\begin{equation*}
{\rm
BIC}_{\lambda_1,\lambda_2}=\log\Big(\frac{S_{\lambda_1,\lambda_2}}{n}\Big)+df_{\lambda_1,\lambda_2}\frac{\log
n}{n},
\end{equation*}
where  
$$
S_{\lambda_1,\lambda_2}=\sum_{i=1}^{n}\sum_{j=1}^{m_{i}}(Y_{ij}-Z_{ij}^{{\rm
T}}\hat{\bm\theta}-\hat{g}(X_{ij}^{{\rm T}}\hat{\bm\beta}))^{2},
$$ 
and $df_{\lambda_1,\lambda_2}$ denotes the number of nonzero components of the estimated parameters.

\section{Numerical examples}\label{sec6}

\subsection{Simulation studies}\label{sec6-1}

In this subsection, we present some simulation studies to evaluate the finite sample performance of the proposed estimation. We denote the bias-corrected GEE estimators as $\hat{\bm \beta}_G, \hat{\bm \theta}_G$, $\hat{g}_G(\cdot)$ and the bias-corrected QIF estimators as $\hat{\bm \beta}_Q, \hat{\bm \theta}_Q$, $\hat{g}_Q(\cdot)$. The working independence estimators $\hat{\bm\beta}_I, \hat{\bm\theta}_I$ and $\hat{g}_I(\cdot)$ are used as comparison in these examples. { In our simulations, we also compare our proposed methods with the method of \cite{wang2009consistent}. The proposed estimators of \cite{li2010empirical} are the same as our bias-corrected GEE estimators. Note that in Wang and Qu's paper parametric models (in particular linear models) are considered. However a linear model does not work well for our simulated data which are generated from a nonlinear model and thus the results are not reported here. Instead, we assume the true $g$ is known and used to compute Wang and Qu's estimator which is denoted by $\hat{\bm \beta}_{WQ}, \hat{\bm \theta}_{WQ}$. } In order to evaluate the variable selection procedure proposed in Section {\ref{sec4}}, the oracle estimators $\hat{\bm\beta}_O$ and $\hat{\bm\theta}_O$ are computed as a comparison, where the zero components are known a priori.

To measure the performance of the proposed estimators, the biases and standard errors of the estimators of ${\bm\beta}_0$ and ${\bm\theta}_0$  are reported. We also define the mean squared errors of the estimators of
${\bm\beta}_0$ and ${\bm\theta}_0$ and $g(\cdot)$ as
$$
{\rm
 MSE}_{\hat{\bm\beta}_{H}}=\frac{1}{L}\sum_{i=1}^L\left(\frac{1}{p}\sum_{k=1}^p(\hat{\beta}_{k,H}^{(i)}-{\beta}_{k})^2\right),\quad
{\rm
MSE}_{\hat{\bm\theta}_{H}}=\frac{1}{L}\sum_{i=1}^L\left(\frac{1}{q}\sum_{k=1}^q(\hat{\theta}_{k,H}^{(i)}-{\theta}_{k})^2\right)
$$
and
$$
{\rm
MSE}_{\hat{g}_{H}}=\frac{1}{L}\sum_{i=1}^L\left(\frac{1}{n}\sum_{j=1}^n\left(\frac{1}{m_j}\sum_{k=1}^{m_i}(\hat{g}_{H}(\hat{t}_k^{(i)})-g(t_k))^2\right)\right),\
\hat{t}_k^{(i)}=X_{jk}^{\rm T}\hat{\bm\beta}_H^{(i)},\
t_k=X_{jk}^{\rm T}\bm\beta_0,
$$
where $H$ denotes $I$, $G$,  $Q$ or $WQ$, and $L$ is the number of replications.  To evaluate the performance of the proposed variable selection method, we used the following cirteria.
\begin{itemize}
  \item The square of the R statistic: $R^2_{\bm\beta}=\frac{|\hat{\bm\beta}^{\rm T}\bm\beta_0|^2}{|\bm\beta_0^{\rm T}\bm\beta_0|^2}$
  and $R^2_{\bm\theta}=\frac{|\hat{\bm\theta}^{\rm T}\bm\theta_0|^2}{|\bm\theta_0^{\rm T}\bm\theta_0|^2}.$
  \item The numbers of zero coefficients and nonzero coefficients obtained by different methods: ``TN'' is the average number of zero coefficients correctly estimated as zero, and ``TP'' is the number of nonzero coefficients identified as nonzero.
\end{itemize}

The data are generated from the following model:
\begin{eqnarray}\label{Sim1}
Y_{ik}=g(t_{ik})+Z_{ik}^{\rm T}\bm\theta_0+e_{ik},\quad
t_{ik}=X_{ik}^{\rm T}\bm\beta_0,
\end{eqnarray}
where $g(t)=e^t$, $X_{ik}=(X_{ik1},\ldots, X_{ikp})^{\rm T}$ and $Z_{ik}$ are generated from $N(0, I_{m_i})$ and $U(0,1)$, respectively, $i=1,\ldots,n; k=1,\ldots,m_i$. The error $\bm e_i=(e_{i1},\ldots, e_{im_i})^{\rm T}$ follows an $m_i$-dimensional multivariate normal distribution with mean 0 and covariance $\Sigma_i=\sigma_i^2\Sigma_0^{m_i},i=1,\ldots,n$. Here we consider two different types of correlation matrix $\Sigma_0^{m_i}$, one is the exchangeable correlation structure and the other is the AR(1) correlation structure with $\rho=0.6$. The kernel function used here is $K(x)=\frac{3}{4}(1-x^{2})$ if $|x|\leq1$, 0 otherwise. The bandwidth is obtained through the leave-one-out cross-validation bandwidth selection method.

\textbf{Example 1}. For model (\ref{Sim1}), let $p=3, q=1, m_1=\cdots=m_n=m=3$, $\bm\beta_0=\frac{1}{\sqrt{14}}(3,2,1)^{\rm T}$ and $\bm\theta_0=0.3$. Let $\sigma_i=1, i=1,\ldots,[n/2]$, and $\sigma_i=2, i=[n/2]+1,\ldots,n$, where $[x]$ denotes the integer part of $x$. The true covariance matrix $\Sigma_0^m$ has an exchangeable correlation structure or an AR(1) correlation structure. The sample size for the simulated data is $n=60$ or $120$, and the number of the simulated datasets is 1000. Two working correlation matrices, exchangeable and AR(1), are considered. We report the results in Tables 1--2.

\begin{table}
\caption{The biases, standard errors and MSE of the proposed estimators for Example 1 with the true correlation matrix being exchangeable (the values in the parentheses are the corresponding standard errors of the estimators).}\label{Table1}
\centering\scriptsize \tabcolsep 0.15cm \begin{tabular}{ccccccccc}
  \hline
 $n$          & Exchangeable&$\beta_1$        & $\beta_2$       & $\beta_3$      &$\theta$       &${\rm MSE}_{\hat{\bm\beta}}$  &${\rm MSE}_{\hat{\theta}}$&${\rm MSE}_{\hat{g}}$ \\
              \hline
 $60$         &Independence& 0.0012(0.0431) & -0.0013(0.0605) & 0.0019(0.0672) & 0.2750(0.7156)&0.0034&0.5873&1.3824 \\
              &GEE    & 0.0017(0.0288) & -0.0023(0.0399) & -0.0005(0.0486)& 0.0756(0.3669)&0.0016&0.1402&1.2189 \\
              &QIF    & 0.0007(0.0316) & -0.0014(0.0436)  & 0.0007(0.0510)& 0.0546(0.3288)&0.0018&0.1110&1.2007 \\
              &QIF$_{WQ}$  & 0.0007(0.0253) & -0.0011(0.0356)  & 0.0006(0.0403)& 0.0052(0.2147)&0.0012&0.0472&--\\
 $120$        &Independence&-0.0008(0.0251) &  0.0009(0.0358) & 0.0007(0.0400) & 0.1462(0.4294)&0.0012&0.2056&0.9342 \\
              &GEE    &-0.0005(0.0173) &  0.0010(0.0244) & -0.0004(0.0267)& 0.0426(0.2181)&0.0005&0.0493&0.8956 \\
              &QIF    &-0.0001(0.0179) &  0.0003(0.0253)  &-0.0002(0.0278)& 0.0331(0.2037)&0.0006&0.0425&0.8862 \\
              &QIF$_{WQ}$  &-0.0002(0.0153) &  0.0005(0.0216)  &-0.0005(0.0267)& 0.0074(0.1436)&0.0005&0.0207&--\\
              \hline
 $n$          & AR(1) & $\beta_1$        & $\beta_2$       & $\beta_3$      &$\theta$       &${\rm MSE}_{\hat{\bm\beta}}$  &${\rm MSE}_{\hat{\theta}}$&${\rm MSE}_{\hat{g}}$ \\
              \hline
 $60$         &Independence& 0.0012(0.0431) & -0.0013(0.0605) & 0.0019(0.0672) & 0.2750(0.7156)&0.0034&0.5873&1.3824 \\
              &GEE    & 0.0016(0.0308) & -0.0017(0.0434) & -0.0015(0.0505)& 0.0714(0.3679)&0.0018&0.1403&1.2196 \\
              &QIF    & 0.0009(0.0331) & -0.0009(0.0505)  &-0.0009(0.0537)& 0.0624(0.3585)&0.0020&0.1323&1.2151 \\
              &QIF$_{WQ}$  &-0.0005(0.0277) & -0.0002(0.0387)  & 0.0007(0.0423)& 0.0068(0.2299)&0.0014&0.0528&--\\
 $120$        &Independence&-0.0008(0.0251) &  0.0009(0.0358) & 0.0007(0.0400) & 0.1426(0.4294)&0.0012&0.2056&0.9342 \\
              &GEE    &-0.0001(0.0179) &  0.0002(0.0254) & -0.0002(0.0289)& 0.0459(0.2265)&0.0006&0.0533&0.8936 \\
              &QIF    & 0.0003(0.0190) & -0.0002(0.0270)  &-0.0004(0.0296)& 0.0334(0.2166)&0.0007&0.0480&0.8958 \\
              &QIF$_{WQ}$  & 0.0003(0.0164) & -0.0001(0.0233)  &-0.0007(0.0267)& 0.0074(0.1529)&0.0005&0.0234&--\\
              \hline
\end{tabular}
\end{table}

\begin{table}
\caption{The biases, standard errors and MSE of the proposed estimators for Example 1 with the true correlation matrix being AR(1) (the values in the parentheses are the corresponding standard errors of the estimators).}\label{Table2} \centering\scriptsize
\tabcolsep 0.15cm \begin{tabular}{ccccccccc}
  \hline
 $n$          & Exchangeable&$\beta_1$        & $\beta_2$       & $\beta_3$      &$\theta$       &${\rm MSE}_{\hat{\bm\beta}}$  &${\rm MSE}_{\hat{\theta}}$&${\rm MSE}_{\hat{g}}$ \\
              \hline
 $60$         &Independence& 0.0011(0.0431) & -0.0012(0.0602) & 0.0020(0.0679) & 0.2708(0.7109)&0.0034&0.5782&1.3829 \\
              &GEE    & 0.0013(0.0303) & -0.0021(0.0417) &  0.0004(0.0518)& 0.0718(0.3644)&0.0018&0.1378&1.2087 \\
              &QIF    & 0.0006(0.0332) & -0.0017(0.0460)  & 0.0017(0.0536)& 0.0558(0.3558)&0.0020&0.1296&1.2087 \\
              &QIF$_{WQ}$  & 0.0007(0.0268) & -0.0011(0.0378)  & 0.0003(0.0450)& 0.0062(0.2219)&0.0014&0.0492&--\\
 $120$        &Independence&-0.0010(0.0250) &  0.0012(0.0355) & 0.0007(0.0397) & 0.1421(0.4289)&0.0012&0.2040&0.9294 \\
              &GEE    &-0.0008(0.0182) &  0.0012(0.0256) & -0.0002(0.0283)& 0.0428(0.2292)&0.0006&0.0543&0.9007 \\
              &QIF    &-0.0004(0.0191) &  0.0007(0.0253)  &-0.0001(0.0295)& 0.0325(0.2163)&0.0006&0.0478&0.8793 \\
              &QIF$_{WQ}$  &-0.0005(0.0164) &  0.0009(0.0233)  &-0.0004(0.0281)& 0.0056(0.1469)&0.0005&0.0216&--\\
              \hline
 $n$       &AR(1) &$\beta_1$        & $\beta_2$       & $\beta_3$      &$\theta$       &${\rm MSE}_{\hat{\bm\beta}}$  &${\rm MSE}_{\hat{\theta}}$&${\rm MSE}_{\hat{g}}$ \\
              \hline
 $60$         &Independence& 0.0011(0.0431) & -0.0012(0.0602) & 0.0020(0.0679) & 0.2708(0.7109)&0.0034&0.5782&1.3829 \\
              &GEE    & 0.0015(0.0302) & -0.0020(0.0422) & -0.0006(0.0493)& 0.0673(0.3509)&0.0017&0.1275&1.1993 \\
              &QIF    & 0.0011(0.0324) & -0.0012(0.0452)  &-0.0008(0.0532)& 0.0594(0.3480)&0.0020&0.1245&1.2185 \\
              &QIF$_{WQ}$  & 0.0001(0.0267) & -0.0003(0.0370)  & 0.0005(0.0416)& 0.0061(0.2192)&0.0013&0.0480&--\\
 $120$        &Independence&-0.0010(0.0250) &  0.0012(0.0355) & 0.0007(0.0397) & 0.1421(0.4289)&0.0012&0.2040&0.9294 \\
              &GEE    &-0.0005(0.0174) &  0.0007(0.0247) & -0.0001(0.0281)& 0.0447(0.2216)&0.0006&0.0510&0.8968 \\
              &QIF    & 0.0000(0.0188) &  0.0007(0.0263)  &-0.0004(0.0289)& 0.0313(0.2079)&0.0006&0.0442&0.8866 \\
              &QIF$_{WQ}$  & 0.0001(0.0160) &  0.0001(0.0227)  &-0.0006(0.0260)& 0.0068(0.1453)&0.0005&0.0212&--\\
              \hline
\end{tabular}
\end{table}

 \textbf{Example 2}. We use the same model as in Example 1, with unbalanced cluster sizes. Let $m_i=3, \sigma_i=1, i=1,\ldots,[n/3]$, $m_i=4, \sigma_i=2, i=[n/3]+1,\ldots,[2n/3]$ and $m_i=5, \sigma_i=3, i=[2n/3]+1,\ldots,n$. We report the results in Tables \ref{Table3}--\ref{Table4}.

\begin{table}
\caption{The biases, standard errors and MSE of the proposed estimators for Example 2 with the true correlation matrix being exchangeable (the values in the parentheses are the corresponding standard errors of the estimators).}\label{Table3}
\centering\scriptsize \tabcolsep 0.15cm \begin{tabular}{ccccccccc}
  \hline
 $n$          & Exchangeable&$\beta_1$        & $\beta_2$       & $\beta_3$      &$\theta$       &${\rm MSE}_{\hat{\bm\beta}}$  &${\rm MSE}_{\hat{\theta}}$&${\rm MSE}_{\hat{g}}$ \\
              \hline
 $60$         &Independence& -0.0021(0.0546) & 0.0016(0.0747) & 0.0029(0.0864) & 0.2357(0.6789)&0.0054&0.5159&1.3987 \\
              &GEE    &-0.0011(0.0394) & 0.0007(0.0549) &  0.0019(0.0621)& 0.0870(0.4032)&0.0028&0.1700&1.3735 \\
              &QIF    &-0.0009(0.0401) & 0.0002(0.0563)  & 0.0021(0.0651)& 0.0738(0.4176)&0.0030&0.1796&1.3799 \\
              &QIF$_{WQ}$  &-0.0012(0.0270) & 0.0008(0.0383)  & 0.0019(0.0446)& 0.0074(0.2480)&0.0014&0.0615&--\\
 $120$        &Independence&-0.0002(0.0325) &  0.0008(0.0454) &-0.0011(0.0523) & 0.1453(0.4500)&0.0020&0.2235&0.9206 \\
              &GEE    &-0.0008(0.0206) &  0.0010(0.0292) & 0.0004(0.0333)& 0.0499(0.2383)&0.0008&0.0592&0.9001 \\
              &QIF    &-0.0005(0.0212) &  0.0008(0.0299)  &-0.0001(0.0345)& 0.0407(0.2234)&0.0008&0.0515&0.9070 \\
              &QIF$_{WQ}$  &-0.0005(0.0157) &  0.0007(0.0230)  & 0.0002(0.0258)& 0.0009(0.0267)&0.0005&0.0267&--\\
              \hline
 $n$          & AR(1) & $\beta_1$        & $\beta_2$       & $\beta_3$      &$\theta$       &${\rm MSE}_{\hat{\bm\beta}}$  &${\rm MSE}_{\hat{\theta}}$&${\rm MSE}_{\hat{g}}$ \\
              \hline
 $60$         &Independence& -0.0021(0.0546) & 0.0016(0.0747) & 0.0029(0.0864) & 0.2357(0.6789)&0.0054&0.5159&1.3987 \\
              &GEE    &-0.0013(0.0418) & 0.0016(0.0574) & 0.00107(0.0655)& 0.0866(0.4150)&0.0031&0.1796&1.3935 \\
              &QIF    &-0.0015(0.0417) & 0.0015(0.0589)  &0.0016(0.0673)& 0.0607(0.4220)&0.0033&0.1816&1.3761 \\
              &QIF$_{WQ}$  &-0.0015(0.0296) & 0.0017(0.0407)  & 0.0011(0.0457)& 0.0085(0.2611)&0.0015&0.0682&--\\
 $120$        &Independence&-0.0002(0.0325) &  0.0008(0.0454) &-0.0011(0.0523) & 0.1453(0.4500)&0.0020&0.2235&0.9206 \\
              &GEE    &-0.0006(0.0217) &  0.0006(0.0312) &  0.0006(0.0349)& 0.0446(0.2541)&0.0009&0.0665&0.8915 \\
              &QIF    &-0.0004(0.0221) &  0.0008(0.0315)  &-0.0004(0.0364)& 0.0289(0.2358)&0.0009&0.0564&0.9028 \\
              &QIF$_{WQ}$  &-0.0005(0.0169) &0.0009(0.0246)  &-0.0004(0.0275)&-0.0007(0.0290)&0.0005&0.0290&--\\
              \hline
\end{tabular}
\end{table}

\begin{table}
\caption{The biases, standard errors and MSE of the proposed estimators for Example 2 with the true correlation matrix being AR(1) (the values in the parentheses are the corresponding standard errors of the estimators).}\label{Table4} \centering\scriptsize
\tabcolsep 0.15cm \begin{tabular}{ccccccccc}
  \hline
 $n$          & Exchangeable&$\beta_1$        & $\beta_2$       & $\beta_3$      &$\theta$       &${\rm MSE}_{\hat{\bm\beta}}$  &${\rm MSE}_{\hat{\theta}}$&${\rm MSE}_{\hat{g}}$ \\
              \hline
 $60$         &Independence&-0.0030(0.0541) & 0.0020(0.0754) & 0.0051(0.0860) & 0.2442(0.6909)&0.0054&0.5366&1.4187 \\
              &GEE    &-0.0014(0.0425) & -0.0005(0.0599) &  0.0051(0.0679)& 0.0782(0.4145)&0.0033&0.1777&1.3623 \\
              &QIF    &-0.0008(0.0434) & -0.0009(0.0614)  & 0.0042(0.0701)& 0.0678(0.4449)&0.0035&0.2023&1.3830 \\
              &QIF$_{WQ}$  &-0.0015(0.0316) & 0.0012(0.0425)  & 0.0021(0.0508)& 0.0062(0.2534)&0.0018&0.0642&--\\
 $120$        &Independence& 0.0005(0.0323) &-0.0000(0.0462) &-0.0014(0.0521) & 0.1463(0.4504)&0.0020&0.2240&0.9068 \\
              &GEE    &-0.0006(0.0226) &  0.0007(0.0316) &  0.0004(0.0362)& 0.0509(0.2553)&0.0009&0.0677&0.9036 \\
              &QIF    &-0.0004(0.0232) &  0.0008(0.0325)  &-0.0002(0.0378)& 0.0421(0.2485)&0.0010&0.0635&0.9052 \\
              &QIF$_{WQ}$  &-0.0006(0.0174) &  0.0008(0.0251)  &0.0002(0.0286)& 0.0004(0.1651)&0.0006&0.0272&--\\
              \hline
 $n$       &AR(1) &$\beta_1$        & $\beta_2$       & $\beta_3$      &$\theta$       &${\rm MSE}_{\hat{\bm\beta}}$  &${\rm MSE}_{\hat{\theta}}$&${\rm MSE}_{\hat{g}}$ \\
              \hline
 $60$         &Independence&-0.0030(0.0541) & 0.0020(0.0754) & 0.0051(0.0860) & 0.2442(0.6909)&0.0054&0.5366&1.4187 \\
              &GEE    &-0.0011(0.0413) & -0.0001(0.0577) &  0.0035(0.0642)& 0.0850(0.4100)&0.0031&0.1751&1.3768 \\
              &QIF    &-0.0016(0.0419) &  0.0006(0.0593)  & 0.0034(0.0676)& 0.0525(0.3982)&0.0033&0.1612&1.3851 \\
              &QIF$_{WQ}$  &-0.0015(0.0282) & 0.0018(0.0400)  & 0.0010(0.0451)& 0.0099(0.2448)&0.0015&0.0600&--\\
 $120$        &Independence& 0.0005(0.0323) &-0.0000(0.0462) &-0.0014(0.0521) & 0.1463(0.4504)&0.0020&0.2240&0.9068 \\
              &GEE    &-0.0005(0.0208) &  0.0006(0.0298) &  0.0003(0.0341)& 0.0504(0.2654)&0.0008&0.0729&0.8981 \\
              &QIF    &-0.0006(0.0216) &  0.0012(0.0307)  &-0.0006(0.0352)& 0.0301(0.2323)&0.0009&0.0548&0.9004 \\
              &QIF$_{WQ}$  & -0.0005(0.0163) &  0.0009(0.0235)  &-0.0002(0.0269)& 0.0006(0.1577)&0.0005&0.0248&--\\
              \hline
\end{tabular}
\end{table}

From Tables \ref{Table1}--\ref{Table4}, by the biases, standard errors and MSE of the proposed estimators, the bias-corrected GEE estimators and bias-corrected QIF estimators have better performance than the working independence estimators. And when the working correlation models are correctly specified, the performances of the estimators are usually slightly better. On the other hand, even when the working correlation is misspecified, both proposals still have comparable performance. Since we use the true $g(\cdot)$ to compute the estimators of \cite{wang2009consistent}, their estimators perform the best because their method avoids estimating the index function. 
When the sample size becomes big, the bias-corrected QIF method is a slightly better choice for estimating the unknown parameters in terms of MSE. For sample size $n=60$, the bias-corrected GEE  and the bias-corrected QIF methods use about 12 seconds for each simulated data set, and for sample size $n=120$, these two methods use about 16 seconds, implemented in Matlab on our PC.

\textbf{Example 3}. For model (\ref{Sim1}), {let $p=20, q=30, m_1=\cdots=m_n=m=3$, $\bm\beta_0=\frac{1}{\sqrt{14}}(3,2,1,{\bf 0}^{\rm T})^{\rm T}$, $\bm\theta_0=(3,1.5,{\bf 0}^{\rm T})^{\rm T}$,} and the covariance matrix $\Sigma_i=\sigma_1^2\Sigma_0^m, \sigma_1=0.5$, $i=1,\ldots,[n/3]$; $\Sigma_i=\sigma_2^2\Sigma_0^m, \sigma_2=1.0$, $i=[3/n]+1,\ldots,[2n/3]$; $\Sigma_i=\sigma_3^2\Sigma_0^m, \sigma_3=2.0$, $i=[2n/3]+1,\ldots,n$. $\Sigma_0^m$ is an exchangeable correlation matrix, and the working correlation matrix is assumed to be exchangeable or AR(1). The sample sizes for the simulated data are $n=50, 100, 200$ and $400$, and the number of simulated data sets is 100. We report the results in Tables \ref{Table5} and \ref{Table6}.

\begin{table}
\caption{ Simulation results for Example 3 with working correlation being exchangeable (the values in the parentheses are the corresponding standard errors). Here ${\bm\xi}^*=(\bm\beta_0^{\rm T},\bm\theta_0^{\rm T})^{\rm T}$.}\label{Table5}\tabcolsep 0.1cm
\centering\begin{tabular}{cccccccc}
  \hline
    &                &       &     & Exchangeable &   &   &\\
   \cline{3-8}
  $n$  & Method          & $R^2_{\bm\beta}$  &  ${\rm TN}_{\bm\beta}$   & ${\rm TP}_{\bm\beta}$  & $R^2_{\bm\theta}$  &  ${\rm TN}_{\bm\theta}$   & ${\rm TP}_{\bm\theta}$   \\ \hline
   50& $\hat{\bm\xi}^*_O$& 0.9992(0.0011) &  17              & 3             & 1.0486(0.1537)  &   28              &  2 \\
     & $\hat{\bm\xi}^*_I$& 0.9470(0.0708) &  16.46           & 2.56        & 0.9924(0.5084)  &   27.37         &  1.72 \\
     & $\hat{\bm\xi}^*_G$  & 0.9612(0.0763) &  16.54         & 2.77        & 1.0095(0.3235)  &   27.78         &  1.91  \\
     & $\hat{\bm\xi}^*_Q$  & 0.9503(0.0930) &  16.53         & 2.65        & 0.9712(0.2733)  &   27.69            & 1.80  \\
     &$\hat{\bm\xi}^*_{WQ}$& 0.9879(0.0238) &  16.81         & 2.91        & 0.9862(0.1188)  &   27.84            & 1.87  \\
     &                 &                &                 &               &                 &                  &    \\
  100& $\hat{\bm\xi}^*_O$& 0.9996(0.0007) &  17              & 3             & 1.0157(0.0865)  &   28              &  2 \\
     & $\hat{\bm\xi}^*_I$& 0.9862(0.0258) &  16.76         & 2.90          & 1.0780(0.2953)  &   27.69         &  1.94 \\
     & $\hat{\bm\xi}^*_G$  & 0.9938(0.0184) &  16.98          & 2.93         & 1.0449(0.1489)  & 27.92             &2 \\
     & $\hat{\bm\xi}^*_Q$  & 0.9916(0.0221) &  16.96         & 2.91          & 1.0080(0.1124)  &  27.89           & 1.97  \\
     &$\hat{\bm\xi}^*_{WQ}$& 0.9992(0.0012) &  17         & 3        & 1.0033(0.0569)  &   28            & 2  \\
     &                 &                &                 &               &                 &                  &   \\
  200& $\hat{\bm\xi}^*_O$& 0.9999(0.0002) &  17              & 3             & 1.0114(0.0838)  &   28              &  2  \\
     & $\hat{\bm\xi}^*_I$& 0.9954(0.0151) &  16.95         & 2.96          & 1.1078(0.2499)  &   27.82          &  2 \\
     & $\hat{\bm\xi}^*_G$  & 0.9995(0.0136) &  17              & 3        & 1.0367(0.0895)  &  28              & 2  \\
     & $\hat{\bm\xi}^*_Q$  & 0.9963(0.0146) &  16.98             &2.96          & 1.0281(0.0709)  &  27.69             & 2  \\
     &$\hat{\bm\xi}^*_{WQ}$& 0.9997(0.0004) &  17         & 3        & 1.0073(0.0357)  &  28            & 2  \\
     &                 &                &                 &               &                 &                  &   \\
  400& $\hat{\bm\xi}^*_O$& 0.9999(0.0001) &  17              & 3             & 1.0119(0.0345)  &   28              &  2  \\
     & $\hat{\bm\xi}^*_I$& 0.9987(0.0007) &  17              & 3        & 1.0770(0.1876)  &   27.85            &  2 \\
     & $\hat{\bm\xi}^*_G$  & 0.9998(0.0002) &  17             & 3        & 1.0105(0.0699)  &   27.8833         &  2  \\
     & $\hat{\bm\xi}^*_Q$  & 0.9993(0.0004) &  17             & 2.9833             & 0.9975(0.0708)  &  27.8667              & 1.9833  \\
     &$\hat{\bm\xi}^*_{WQ}$& 0.9999(0.0001) &  17         & 3        & 1.0008(0.0255)  &   28            & 2  \\
  \hline
\end{tabular}
\end{table}

\begin{table}
\caption{ Simulation results for Example 3 with working correlation being AR(1) (the values in the parentheses are the corresponding standard errors). Here ${\bm\xi}^*=(\bm\beta_0^{\rm T},\bm\theta_0^{\rm T})^{\rm T}$.}\label{Table6}\tabcolsep 0.1cm
\centering\begin{tabular}{cccccccc}
  \hline
    &                &       &     & AR(1) &   &   &\\
   \cline{3-8}
  $n$  & Method          & $R^2_{\bm\beta}$  &  ${\rm TN}_{\bm\beta}$   & ${\rm TP}_{\bm\beta}$  & $R^2_{\bm\theta}$  &  ${\rm TN}_{\bm\theta}$   & ${\rm TP}_{\bm\theta}$   \\ \hline
   50& $\hat{\bm\xi}^*_O$& 0.9991(0.0012) &  17              & 3             & 1.0383(0.1359)  &  28              &  2 \\
     & $\hat{\bm\xi}^*_I$& 0.9470(0.0708) &  16.46           & 2.65        & 0.9924(0.5084)  &   27.37         &  1.72 \\
     & $\hat{\bm\xi}^*_G$  & 0.9545(0.0927) &  16.51           & 2.76         & 0.9808(0.2268)  &  27.68         & 1.91  \\
     & $\hat{\bm\xi}^*_Q$  & 0.9311(0.1048) &  16.46         & 2.58          & 0.9532(0.3813)  &   27.6          & 1.68  \\
     &$\hat{\bm\xi}^*_{WQ}$& 0.9852(0.0306) &  16.76         & 2.90        & 1.0072(0.1104)  &   27.69            & 1.91  \\
     &                 &                &                 &               &                 &                  &    \\
  100& $\hat{\bm\xi}^*_O$& 0.9995(0.0007) &  17              & 3             & 1.0173(0.0860)  &  28              &  2 \\
     & $\hat{\bm\xi}^*_I$& 0.9862(0.0258) &  16.76         & 2.90          & 1.0780(0.2953)  &   27.69         &  1.94 \\
     & $\hat{\bm\xi}^*_G$  & 0.9909(0.0227) &  16.95         & 2.90          & 1.0342(0.1278)  &   27.76          &2 \\
     & $\hat{\bm\xi}^*_Q$  & 0.9882(0.0347) &  16.91         & 2.90        & 1.0035(0.1743)  &   27.72           & 1.95  \\
     &$\hat{\bm\xi}^*_{WQ}$& 0.9990(0.0013) &  16.99         & 3        & 1.0041(0.0634)  &   28            & 2  \\
     &                 &                &                 &               &                 &                  &   \\
  200& $\hat{\bm\xi}^*_O$& 0.9999(0.0002) &  17             & 3             & 1.0108(0.0787)  &   28             &  2  \\
     & $\hat{\bm\xi}^*_I$& 0.9954(0.0151) &  16.99         & 2.96          & 1.1078(0.2499)  &   27.82          &  2 \\
     & $\hat{\bm\xi}^*_G$  & 0.9991(0.0034) &  17         & 3             & 1.0321(0.0846)  &  27.97           & 1.99  \\
     & $\hat{\bm\xi}^*_Q$  & 0.9965(0.0142) &  17              & 2.96        & 1.0243(0.0960)  &  27.86          & 1.99  \\
     &$\hat{\bm\xi}^*_{WQ}$& 0.9996(0.0004) &  17         & 3        & 1.0073(0.0375)  &   28            & 2 \\
     &                 &                &                 &               &                 &                  &   \\
  400& $\hat{\bm\xi}^*_O$& 0.9999(0.0001) &  17              & 3             & 1.0128(0.0359)  &  28              &  2  \\
     & $\hat{\bm\xi}^*_I$& 0.9993(0.0007) &  17              & 3        & 1.0770(0.1876)  &   27.8833            &  2 \\
     & $\hat{\bm\xi}^*_G$  & 0.9998(0.0002) &  17            & 3             & 1.0119(0.1034)  &  27.8167              &  2  \\
     & $\hat{\bm\xi}^*_Q$  & 0.9998(0.0002) &  17             & 3             & 0.9901(0.0886)  &  28              &  1.9667  \\
     &$\hat{\bm\xi}^*_{WQ}$& 0.9999(0.0001) &  17         & 3        & 1.0005(0.0256)  &   27.9833            & 2  \\
  \hline
\end{tabular}
\end{table}

From Tables \ref{Table5}--\ref{Table6}, we can explore the performance of the penalized bias-corrected GEE estimators and the penalized bias-corrected QIF estimators. The oracle estimators give the perfect values of ${\rm TN}_{\bm\beta}$, ${\rm TP}_{\bm\beta}$, ${\rm TN}_{\bm\theta}$, ${\rm TP}_{\bm\theta}$. From Tables \ref{Table5} and \ref{Table6}, it can be observed that the proposed estimators are close to the oracle estimators in terms of $R^2_{\bm\beta}$ and $ R^2_{\bm\theta}$, which are close to 1. Generally, with the sample size increasing, the proposed method's performance in terms of $R^2_{\bm\beta}$, $ R^2_{\bm\theta}$, ${\rm TN}_{\bm\beta}$, ${\rm TP}_{\bm\beta}$, ${\rm TN}_{\bm\theta}$ and ${\rm TP}_{\bm\theta}$ improves. From Tables \ref{Table5}--\ref{Table6}, it is easy to see that the proposed estimators perform better  in terms of $R^2_{\bm\beta}$ and $ R^2_{\bm\theta}$ when the working correlation structure is correctly specified. In addition, even when the working correlation is misspecified, the bias-corrected penalized GEE method and the bias-corrected penalized QIF method still can identify the important variables. It shows that the proposed penalized methods are not sensitive to the choice of the working correlation structure. In terms of computation time, for each simulated data set, the penalized methods take about 2 minutes, 4 minutes, 12 minutes and 40 minutes for sample size $n=50, 100, 200$ and 400, respectively


To conform further the effect of correlation matrix on variable selection, we now use  AR(1) correlation matrix as the truth and consider bias-corrected penalized GEE estimators using  AR(1) and exchangeable correlation matrix in model fitting. The results are reported in Table \ref{Table77}. The observations we can make are similar as before. It shows that the proposed bias-corrected penalized GEE method is not quite sensitive to the choice of the working correlation structure.

\begin{table}
\caption{ Simulation results for GEE estimator for Example 3 when the true correlation matrix is AR(1) and the working correlation is AR(1) or exchangeable (the values in the parentheses are the corresponding standard errors). Here ${\bf\xi}^*=(\bf\beta_0^{\rm T},\bf\theta_0^{\rm T})^{\rm T}$.}\label{Table77}\tabcolsep 0.1cm
\centering\begin{tabular}{cccccccc}
  \hline
    &                &       &     & AR(1) &   &   &\\
   \cline{3-8}
  $n$  & Method          & $R^2_{\bf\beta}$  &  ${\rm TN}_{\bf\beta}$   & ${\rm TP}_{\bf\beta}$  & $R^2_{\bf\theta}$  &  ${\rm TN}_{\bf\theta}$   & ${\rm TP}_{\bf\theta}$   \\ \hline
   50& $\hat{\bf\xi}^*_G$  & 0.9668(0.0535) &  16.46         & 2.8        & 1.0589(0.4369)  &   27.52         &  1.92  \\
  100& $\hat{\bf\xi}^*_G$  & 0.9948(0.0174) &  16.98          & 2.95         & 1.0540(0.1633)  & 27.76             &2 \\
  200& $\hat{\bf\xi}^*_G$  & 0.9995(0.0007) &  17              & 3        & 1.0490(0.1398)  &  27.96              & 2  \\
  400& $\hat{\bf\xi}^*_G$  & 0.9998(0.0002) &  17             & 3        & 1.0189(0.0804)  &   28         &  2  \\
  \hline
    &                &       &     & Exchangeable &   &   &\\
   \cline{3-8}
   50& $\hat{\bf\xi}^*_G$  & 0.9657(0.0608) &  16.45           & 2.75         & 1.0692(0.4827)  &  27.35         & 1.96  \\
  100& $\hat{\bf\xi}^*_G$  & 0.9901(0.0294) &  16.95         & 2.92          & 1.0386(0.1729)  &   27.66          &1.99 \\
  200& $\hat{\bf\xi}^*_G$  & 0.9983(0.0115) &  17         & 3             & 1.0317(0.1269)  &  27.96           & 2  \\
  400& $\hat{\bf\xi}^*_G$  & 0.9998(0.0002) &  17            & 3             & 1.0194(0.1083)  &  27.8375              &  2  \\
     \hline
\end{tabular}
\end{table}

\vskip0.2cm

\subsection{Application to CD4 data}\label{sec6-2}

We now apply the method to the CD4 data from the Multi-Center AIDS Cohort Study. This data set consists of 283 homosexual males who were HIV positive between 1984 and 1991. All individuals were scheduled to have their measurements taken during semiannual visits. Each patient had a different number of repeated measurements and the true observation times were not equally spaced because patients often missed or rescheduled their appointments. Details of the study were described in \cite{huang2002varying} and \cite{fan2004new}. \cite{qu2006quadratic} analyzed the same data set using varying coefficient models. Here we apply the partially linear single-index model to this data set.

\begin{figure}[ptbh]
\setlength{\parindent}{-2.5em} \centering
\scalebox{0.45}{\includegraphics{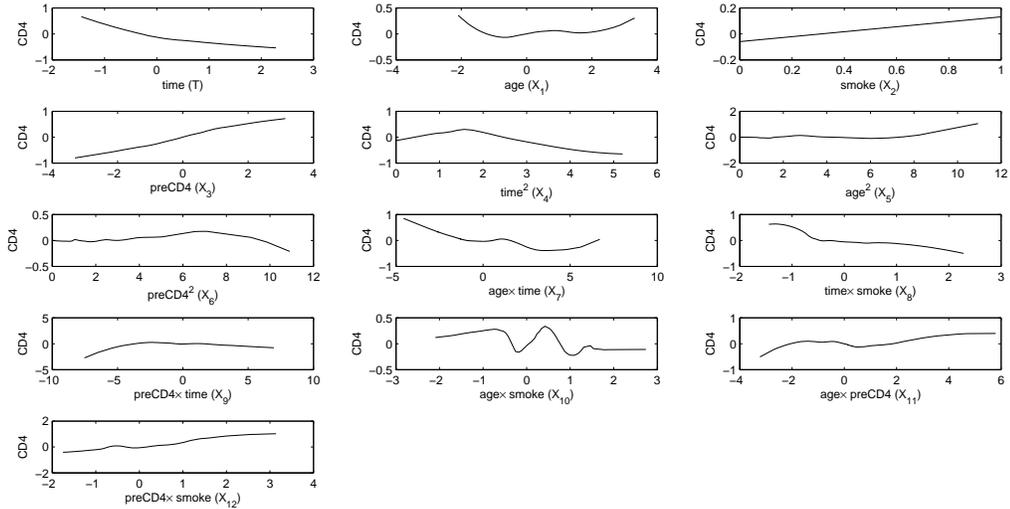}} \caption{The kernel smoothing plots between each covariate and response variable.\label{fig-2}}
\end{figure}

\begin{figure}[ptbh]
\setlength{\parindent}{-2.5em} \centering
\scalebox{0.6}{\includegraphics{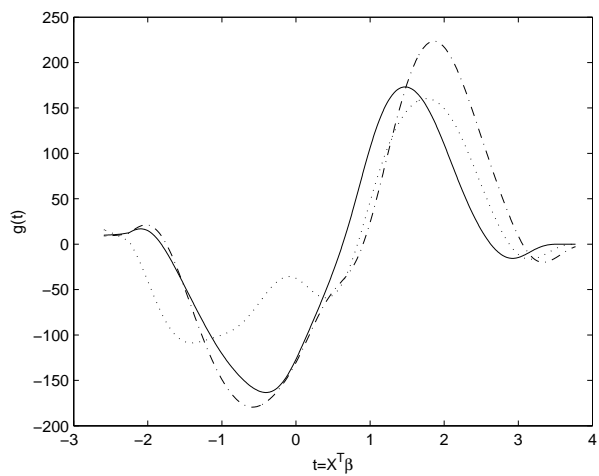}} \caption{The estimated link function $g(t)$ where the solid line is obtained by GEE, the dotted line is obtained by QIF and the dashed line is obtained by the independence working correlation matrix. \label{fig-3}}
\end{figure}

The response variable $Y$ is the CD4 percentage over time. Four covariates are also collected: $T$, patients' visiting time; $X_1$, patient's age; $X_2$, the individual's smoking status, which takes binary values 1 or 0, according to whether a individual is a smoker or nonsmoker; $X_3$, the CD4 cell percentage before infection. We also consider the squares and cross multiples of these covariates, which include $X_4=T^2, X_5=X_1^2, X_6=X_3^2, X_7=T\times X_1, X_8=T\times X_2, X_9=T\times X_3, X_{10}=X_1\times X_2, X_{11}=X_1\times X_3, X_{12}=X_2\times X_3$. In order to apply the partially linear single-index model, we first divide all these covariates into two groups, the linear part and the single-index part as follows. After standardizing these variables, the kernel smoothing plots between each covariate and the response variable are shown in Figure \ref{fig-2}. From visual inspection we use variables $Z=(T, X_2, X_3, X_5, X_9, X_{12})^{\rm T}$ in the linear part, and the others in the single-index part. We compare our method with that of \cite{wang2009consistent}, for which we assume the model is linear ($g$ is a linear function). We apply the bias-corrected penalized GEE procedure, the bias-corrected penalized QIF procedure, and the procedure proposed by \cite{wang2009consistent} with the AR(1) working correlation matrix. And we also consider the bias-corrected penalized GEE procedure with independent working correlation matrix. Applying these procedures, we can select the important variables starting from the following model$$
{\bm Y}_{i}={\bm Z}_{i}{\bm\theta}+g({\bm X}_{i}{\bm\beta})+{\bm
e}_{i},
$$
where
$$
\bm Y_i=\left(
\begin{array}{c}
      Y_{i1} \\
      \vdots \\
      Y_{m_i}
    \end{array}
        \right),\quad
\bm Z_i=\left(
      \begin{array}{cccccc}
      T_{i_1} & X_{i2_1} & X_{i3_1} & X_{i5_1} & X_{i9_1} & X_{i12_1}\\
       \cdots & \cdots &  \cdots &  \cdots &  \cdots &  \cdots \\
       T_{i_{m_i}} & X_{i2_{m_i}} & X_{i3_{m_i}}  & X_{i5_{m_i}} & X_{i9_{m_i}} & X_{i12_{m_i}} \\
      \end{array}
    \right),
$$
$$
\bm X_i=\left(
      \begin{array}{ccccccc}
      X_{i1_1} & X_{i4_1} & X_{i6_1} & X_{i7_1} & X_{i8_1} & X_{i10_1} & X_{i11_1} \\
       \cdots & \cdots &  \cdots &  \cdots &  \cdots &  \cdots &  \cdots  \\
       X_{i1_{m_i}} & X_{i4_{m_i}} & X_{i6_{m_i}} & X_{i7_{m_i}} & X_{i8_{m_i}} & X_{i10_{m_i}} & X_{i11_{m_i}} \\
      \end{array}
    \right),\quad
\bm e_i=\left(
\begin{array}{c}
      \varepsilon_{i1} \\
      \vdots \\
      \varepsilon_{m_i}
    \end{array}
        \right).
    $$
The estimated nonzero parameters and their $95\%$ confidence intervals are reported in Table \ref{Table7}. Here, these intervals are constructed using similar methods as in \cite{carroll1997generalized}. In order to compare the mean squared prediction error (MSPE), we use five-fold cross-validation, and the results are also shown in Table \ref{Table7}. From Table \ref{Table7}, we see that all the methods identify similar models, and the bias-corrected penalized GEE procedure has the best performance based on MSPE although the differences among various methods are small. Wang and Qu's method has the largest MSPE suggesting the assumption of $g(\cdot)$ being linear is probably wrong. Furthermore, if we focus on the variable selection problem, the bias-corrected penalized QIF method obtains the smallest numbers of significant variables with similar MSPE to the bias-corrected penalized GEE method. 
The fitted curves for the unknown link function $g(\cdot)$ are shown in Figure \ref{fig-3}.

\begin{table}
\caption{Estimates and confidence intervals of the nonzero parameters for the real data.}\label{Table7}
\tabcolsep
0.05cm\vskip0.3cm{\small \centering\begin{tabular}{ccccccc|c}
  \hline
    &      &Estimates & Confidence interval &             &Estimates  & Confidence interval&${\rm MSPE}$                            \\
              \hline
GEE&$\beta_{X_1}$ &-0.6627 & [-0.7469,-0.5785]  & $\beta_{X_4}$ & 0.6124 & [0.5354,0.6894]&0.7294\\
&$\beta_{X_7}$ & -0.4310 & [-0.4703,-0.3918]      &   &  &  &\\
&$\theta_{T}$  &-0.3528 & [-0.4617,-0.2439]     & $\theta_{X_3}$
&0.4527 & [0.2535,0.6519]&\\ \hline
QIF&$\beta_{X_1}$ &-0.9212 & [-0.9674,-0.8751]  & $\beta_{X_4}$ & 0.3891 & [0.2797,0.4984]&0.7330\\
&$\theta_{T}$  &-0.3789 & [-0.5263,-0.2314]     & $\theta_{X_3}$
&0.2814& [0.1776,0.3852]&\\ \hline
QIF$_{WQ}$&$\beta_{X_8}$ &-0.1876 & [-0.3865,0.0113]  &  &  & &0.8304\\
&$\theta_{T}$  &-0.3789 & [-0.5263,-0.2314]     & $\theta_{X_3}$
&0.2814& [0.1776,0.3852]&\\ \hline
Independence&$\beta_{X_1}$ &-0.6315 & [-0.6887,-0.5743]      & $\beta_{X_4}$ &0.7754   & [0.7288,0.8220]&0.7332\\
&$\theta_{T}$     &-0.3517  & [-0.4354,-0.2680]     & $\theta_{X_3}$ &0.2824  & [0.1891,0.3757]&\\
  \hline
\end{tabular}}
\end{table}

\section{Concluding remarks}\label{sec7}
In this paper, we proposed the bias-corrected GEE estimator and the bias-corrected QIF estimator for partially linear single-index models with longitudinal data. By taking into account the correlation within each subject, we can improve the performance of the estimators. In addition, variable selection and estimation can be performed at the same time based on penalization. The resulting estimators are consistent in identifying the true model and enjoy the oracle property.

The working correlation structure is used to improve the performance of the estimators. When the working correlation structure is correctly specified,  the proposed estimators perform better than the estimators with the misspecified  working correlation structure. 
When the inverse working correlation matrix can be approximated by a linear combination of several basis matrices, the bias-corrected QIF method can avoid estimating the nuisance parameters in the working correlation matrix. Therefore, it is easier to choose the working correlation structure using the bias-corrected QIF method.

In the real data analysis, we used a heuristic method to separate the predictors into the linear part and the  single-index part. How to separate the predictors into the linear part and the single-index part in a more principled way is an important problem. \cite{zhang2007tests} proposed the generalized likelihood ratio (GLR) statistic to test whether some predictors should be in the linear part. \cite{li2013scb-sim} proposed an adaptive Neyman test statistic to determine which predictors belong to the linear part. Automatic structure identification for single-index models based on penalization, following the recent work of \cite{helen11}, is another interesting direction for future investigation. 


For longitudinal data it is essential to estimate and select a working correlation structure since correctly modeling correlation structure will increase the efficiency of the regression parameter estimator. Estimation and selection of the working correlation structure is a challenging problem. For  linear models and generalized linear models with longitudinal data, some approaches for estimating or selecting a working correlation structure have been proposed. For example, \cite{chen2012el} proposed an empirical likelihood approach to select the best working correlation structure in GEE, \cite{zhou2012scs} proposed an approach to estimate and select the working correlation structure simultaneously through a group penalty strategy, and \cite{pan2001aic,pan2002gee} proposed semiparametric and nonparametric approaches to select the working correlation structure in GEE. Estimation and selection of the working correlation structure in our context is an interesting topic for future research.

\section*{Acknowledgments}

The authors would like to thank the Editor, Associate Editor and two referees for insightful comments that led to an improvement of an earlier manuscript. Gaorong Li's research was supported by NSFC (11101014), the Specialized Research Fund for the Doctoral Program of Higher Education of China (20101103120016), PHR(IHLB, PHR20110822), the Science and Technology Project of Beijing Municipal Education Commission (KM201410005010) and the Fundamental Research Foundation of Beijing University of Technology (X4006013201101). Peng Lai's research was supported by NSFC (11301279, 11226222), Natural Science Foundation of the Jiangsu Higher Education Institutions of China (12KJB110016) and the University Foundation of Nanjing University of Information Science and Technology (No. 20110389).

\vskip 3mm

\bibliographystyle{jasa}

\bibliography{papers}


\end{document}